\documentclass[twocolumn,nobibnotes,showpacs,preprintnumbers,amsmath,amssymb]{revtex4}
\pdfoutput=1
\usepackage{graphicx}% Include figure files
\usepackage{dcolumn}% Align table columns on decimal point
\usepackage{bm}% bold math
\usepackage{multirow}%table formatting

\newcommand{\be}{\begin{equation}}
\newcommand{\ee}{\end{equation}}
\newcommand{\bea}{\begin{eqnarray}}
\newcommand{\eea}{\end{eqnarray}}

\newcommand{\keV}{\;{~\rm keV}}
\newcommand{\keVee}{\;{~\rm keV$_\text{ee}$}}
\newcommand{\keVnr}{\;{~\rm keV$_\text{nr}$}}

\newcommand{\GeV}{\;{~\rm GeV}}

\newcommand{\eqnref}[1]{Eq.~\ref{#1}}
\newcommand{\figref}[1]{Fig.~\ref{#1}}
\newcommand{\vperpinel}{\vec{v}^\perp_\text{inel}}
\newcommand{\qvec}{\vec{q}\,}
\newcommand{\kgday}{kg$\cdot$days}
\newcommand{\vinelsq}{|\vec v_{\text{inel}\,T}^\perp|^2}
\newcommand{\qsqmN}{\frac{|\vec q\,|^2}{m_N^2}}

\newcommand{\vinsqExp}{\left(|\vec v_T\,|^2-v_{\text{min}\,T}^2(\delta)\right)}

\begin{document}
\title{A Model Independent Approach to Inelastic Dark Matter Scattering}
\author{G.~Barello, Spencer Chang, and Christopher A.~Newby}

\affiliation{Department of Physics and Institute of Theoretical Science, University of Oregon, Eugene, Oregon 97403}
\begin{abstract}
We present a model independent analysis of inelastic dark matter transitions at direct detection experiments by modifying the elastic methodology of Fitzpatrick, et al. By analyzing the kinematics of inelastic transitions, we find the relevant variables to describe these scattering processes, the primary change being a modification of the $\vec{v}^\perp$ variable.  Taking this into account, we list the relevant scattering matrix elements and modify the Mathematica package of Anand, et al.\ to calculate the necessary form factors.  As an application, we determine the matrix elements of inelastic scattering for spin transitions between a fermion to fermion, scalar to vector, and scalar to scalar.  Finally, we consider fits to the DAMA/LIBRA annual modulation signal for the magnetic inelastic dark matter scenario as well as a model independent scan over relativistic operators, constraining them with limits from direct detection experiments.  In the magnetic inelastic dark matter scenario or if the dark matter couples through relativistic operators involving only  protons, we find that experiments with xenon  and germanium  targets can have consistently small rates.  However, limits from iodine experiments  are much more constraining, leaving small regions of allowed parameter space.  We point out that existing uncertainties in the iodine quenching factor strongly affects the constraints, motivating further study to pin down the correct values.   
\end{abstract}

\maketitle

\section{Introduction}

Dark matter direct detection experiments are an ambitious effort to observe galactic dark matter scattering off of nuclear targets \cite{Goodman:1984dc} as a means to study dark matter's interactions with normal matter.   Beginning with the early experiments in the eighties, there has been steady progress to increasing sensitivity.  Planned experiments in the future will push this frontier  \cite{Cushman:2013zza}, giving us hope that such interactions will be confirmed soon.  Such a discovery would give important insights into  the fundamental nature of dark matter and its place in the Standard Model of particle physics.  

The experimental challenges of direct detection are many.  Finding conclusive evidence is a tall order, as demonstrated by several recent experimental anomalies, the most famous being the annual modulation signal seen by DAMA \cite{Bernabei:2013xsa},  which appear to be in conflict with the null results of other experiments.  However, whether a dark matter scenario is consistent with existing limits and excesses depends strongly on the form of its interactions with the nucleus.    For each interaction, the relative sensitivities of different experiments can vary wildly, leading to the hope of a scenario consistent with all of the existing data.   Another reason to study the allowed interactions is that certain interactions may have distinctive features in the signal that allow better background separation.    These reasons highlight the importance of exploring the full landscape of possible interactions.   Some examples of the studied possibilities include  inelastic transitions \cite{TuckerSmith:2001hy}, dark matter form factors \cite{Chang:2009yt, Feldstein:2009tr}, dark matter-nucleus resonances \cite{Bai:2009cd}, and isospin-violating dark matter \cite{Chang:2010yk, Feng:2011vu, Cirigliano:2013zta}. 
    
Given the large range of possible scattering interactions allowed by dark matter theories, it has proven useful to study the  phenomenology of dark matter scattering  in a model independent fashion \cite{Fan:2010gt, Fitzpatrick:2012ix}.  In particular, Ref.\ \cite{Fitzpatrick:2012ix} has provided a systematic study of the effective description of nonrelativistic, elastic scattering and a Mathematica package to generate the necessary form factors \cite{Anand:2013yka}.    A notable success of this approach was the discovery of nuclear responses beyond the standard spin-independent and spin-dependent responses that are currently analyzed by experiments.  Thus, model independent approaches have the benefit of larger applicability, pointing out all of the regions where experiments can be sensitive --- see \cite{Fitzpatrick:2012ib, DelNobile:2013sia, Gresham:2014vja, Catena:2014uqa, Gluscevic:2014vga} for some recent work in this direction.  

In this paper, we extend this work by considering the modifications necessary to describe inelastic transitions of the dark matter particle.  Such transitions have important kinematic effects and were originally proposed and studied for scattering to a heavier dark matter state  \cite{TuckerSmith:2001hy, TuckerSmith:2004jv} and then later extended to the ``down scattering" case \cite{Finkbeiner:2009mi, Essig:2010ye, Graham:2010ca}.  We will investigate the modifications to Ref.~\cite{Fitzpatrick:2012ix} that must be made to properly treat inelastic scattering in a model-independent fashion.  As we will show, this requires a straightforward reorganization of the basis of scattering matrix-elements.  This has the added benefit that we were able to suitably modify the Mathematica package \cite{Anand:2013yka}  to calculate the form factors for inelastic scattering.      

To illustrate the utility of this methodology, we will demonstrate how the inelastic transitions between particles of spin 1/2 to 1/2, 0 to 1, and 0 to 0 can be treated in a standard basis of nonrelativistic matrix elements.  We do so by considering the relativistic operators between such particles that can be mediated by spin 0 or 1 particles.  Using these results, we  perform a reanalysis of the magnetic inelastic dark matter scenario \cite{Chang:2010en} and perform a model independent scan over the relativistic operators  to determine scenarios which could explain the DAMA/LIBRA signal.  For the magnetic inelastic dark matter scenario and for operators which couple the dark matter to protons only, we find the constraints from xenon detectors can be weakened to allow some operators to survive, while germanium detectors have an extremely weak sensitivity.   However, a stringent constraint comes from iodine targets, like those used by COUPP and KIMS.  A large uncertainty in this analysis is the quenching factor of iodine.  Depending on the values we assume, the constraints from KIMS, XENON, and LUX can change by a large amount, due to changes in the recoil spectra.  Another uncertainty is the lack of form factors for cesium and tungsten.  Given these uncertainties, we find that DAMA explanations are constrained but not ruled out yet, which should be resolved by the next round of experimental releases.              

The outline of the rest of the paper is as follows.  In section \ref{sec:kinematics}, we discuss the kinematics of inelastic scattering to determine the relevant kinematic variables.  In section \ref{sec:operators} we discuss the modifications to the operators needed to describe dark matter inelastic transitions.  In section \ref{sec:dama}, as an application of this formalism, we fit the annual modulation signal at DAMA/LIBRA and discuss the constraints from other experiments.  In section \ref{sec:conclusions}, we conclude.  Finally, in the appendices, we give further details on the nonrelativistic limit of the kinematics and matrix elements of inelastic scattering.    
\begin{figure}[t]
%\begin{center}
\includegraphics[scale=0.4]{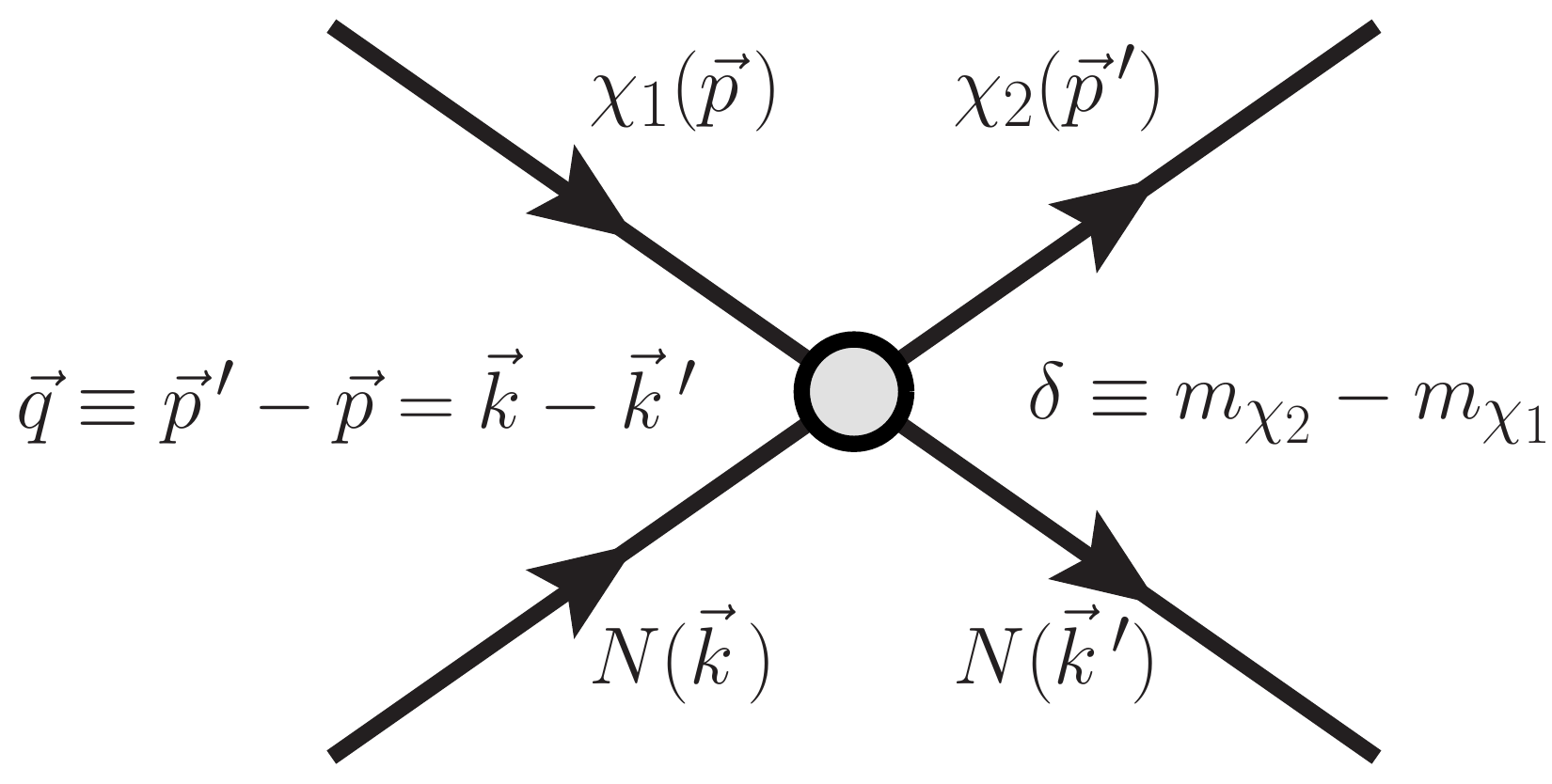}\hspace{.4in}
%\end{center}
%\begin{minipage}{5.5in}
\caption{\label{fig:scattering} Inelastic scattering of dark matter off of a nucleon with our conventions for the kinematic variables.  }
%\end{minipage}
\end{figure}

\section{Variables for Inelastic Kinematics\label{sec:kinematics}}
To begin, we need to determine the correct variables to describe inelastic scattering.  To do so, we need to understand the kinematic modifications of an inelastic transition for nonrelativistic scattering.  We are interested in scattering events of the type
\bea
\chi_1 (\vec{p}\,)\; N (\vec{k}\,) \to \chi_2  (\vec{p}\,')\; N  (\vec{k}\,')\;
\eea
where $\chi_1$ is the incoming dark matter particle, $\chi_2$ is the outgoing particle, and $N$ is a nucleon in the target nucleus, see \figref{fig:scattering}.  There is a mass splitting between the two particles $\delta = m_{\chi_2}-m_{\chi_1}$.  Positive $\delta$ was the first case to be considered originally \cite{TuckerSmith:2001hy}, which pointed out that this has the important effects of favoring scattering off of heavier nuclei and increasing the annual modulation fraction.  Negative $\delta$ leads to exothermic transitions which have also been considered in the literature \cite{Finkbeiner:2009mi, Essig:2010ye, Graham:2010ca}.  In certain theories, the elastic scattering process is forbidden or suppressed \cite{Han:1997wn, Hall:1997ah}, making these inelastic transitions the leading way to detect dark matter scattering.  For a survey of such theories, see \cite{TuckerSmith:2001hy, Cui:2009xq, Alves:2009nf, Kaplan:2009de, Kumar:2011iy}.

The modifications of a nonzero splitting $\delta$ on the kinematics is straightforward.  To leading order in the nonrelativistic expansion, $\delta$ is the additional energy required to make the transition occur.  Thus, given the scaling of kinetic energy, we expect situations where the splitting scales as $\delta \sim O(v^2)$ to have a consistent velocity expansion.  Since dark matter in our galaxy have speeds $v\sim 10^{-3}c$, this means that we should consider splittings in the range $\delta \sim 100 \keV \left(\frac{m_\chi}{100 \GeV}\right)$.  

Now, we adapt the analysis of \cite{Fitzpatrick:2012ix} to inelastic scattering in order to determine the relevant degrees of freedom that characterize the effective field theory in a velocity expansion.  One approach would be to start with the relativistic kinematics and take the nonrelativistic limit.    Although this gives the same result, as we show in Appendix~\ref{app:relkinematics}, we find that it is simpler to proceed from the constraints of Galilean invariance where velocities receive a common shift.   This determines that there are two relevant vectors that are boost invariant, $\vec{v}\equiv \vec{v}_{\chi_1} - \vec{v}_{N_{in}}=\vec{p}/m_{\chi_1}-\vec{k}/{m_N}$ and $\qvec= \vec{p}\,' - \vec{p} = \vec{k}-\vec{k}'$, while the boost invariant scalars are the particle masses and $\delta$.  Note that $\vec{p}\,'-\vec{p}$ is not exactly Galilean invariant; due to the mass difference $\delta$, it is invariant to leading order in the velocity expansion and thus is a consistent approximation at first order.  Throughout this discussion, we are working in this expansion and will cavalierly use equalities for expressions if they are equal to the same order in the expansion.    

At this point,  it is useful to construct an orthogonal basis of these vectors.  To do so, consider the scattering in the center-of-mass frame, where  $\vec{v}_{\chi_1} = \frac{\mu_N}{m_{\chi_1}}\vec{v}$, $\vec{v}_{N_{in}} = -\frac{\mu_N}{m_N}\vec{v}$, and  $\mu_N$ is the reduced mass between $\chi_1$ and $N$.  The initial energy in this frame, expanded to second order in velocities, is
\bea
E_{in} \approx m_{\chi_1} + m_N + \frac{1}{2} \mu_N v^2.  
\eea
After scattering, the momentum vectors are $\vec{p}\,'=\vec{p}+\vec{q}$ and $\vec{k}'=\vec{k}-\vec{q}$.  Expanding the final energy to the same order, we find 
\bea
E_{out}&= &m_{\chi_2}+m_N + \frac{1}{2m_{\chi_2}} |\vec{p}+\qvec|^2+\frac{1}{2m_{N}} |\vec{k}-\qvec|^2 \nonumber \\[-.2cm] \\ [-.2cm]
&\approx & E_{in} +\delta + \vec{v}\cdot \vec{q} + \frac{|\qvec|^2}{2\mu_N}.\nonumber       
\eea
To reach the final form, we treated all momenta as order $v$ and $\delta$ as order $v^2.$  Thus, we find that energy conservation requires 
\bea
\delta + \vec{v}\cdot \vec{q} + \frac{|\vec q\,|^2}{2\mu_N} = 0. \label{eqn:econv}
\eea
Using this constraint, one can easily show that
\bea
\vec{v}^\perp_\text{inel} \equiv \vec{v} + \frac{\vec{q}}{2\mu_N} + \frac{\delta}{|\qvec|^2} \vec{q} = \vec{v}^\perp_\text{el}+ \frac{\delta}{|\qvec|^2} \vec{q}
\eea   
is perpendicular to $\vec{q}$.  Here we see that the inelastic kinematics alters this vector from the elastic version $\vec{v}^\perp_\text{el}$ by a new piece proportional to $\delta$.  This new term is  entirely consistent with the velocity expansion.  

As a consistency check, notice that \eqnref{eqn:econv} requires
\bea
|\vec{v}| \ge \frac{1}{|\vec{q}\,|} \left|\frac{|\qvec|^2}{2\mu_N} + \delta\right|.
\eea
If we write the momentum transfer in terms of the energy recoil $|\vec{q}\,| = \sqrt{2m_N E_R}$, we find that the minimum velocity for scattering is
\bea
v_\text{min} = \frac{1}{\sqrt{2m_N E_R}}\left|\frac{m_N E_R}{\mu_N}+\delta\right|
\label{eqn:minv}
\eea
which reproduces the well known result in the literature \cite{TuckerSmith:2001hy}.  

\section{Inelastic Scattering Operators \label{sec:operators}}
Now that we know the correct variables to describe inelastic kinematics, we can now list the allowed matrix elements for inelastic, nonrelativistic dark matter-nucleon scattering.  To leading order in the velocity expansion, we found that the only modification is that $\vec{v}^\perp$ is changed from the elastic case.  Thus, the operators that are allowed are the same as in \cite{Fitzpatrick:2012ix} with $\vec{v}^\perp \to \vperpinel.$  Listing these in the same numbering scheme, we have
\be
\label{eqn:operators}
\begin{split}
&\mathcal{O}_1 ={\bf 1}_\chi{\bf 1}_N, \quad \mathcal{O}_2=(v_\text{inel}^\perp)^2,\quad \mathcal{O}_3 =i \vec S_N\cdot\left(\frac{\vec q}{m_N}\times\vec v_\text{inel}^\perp\right),\\
&\mathcal{O}_4=\vec S_\chi\cdot\vec S_N,\quad \mathcal{O}_5=i \vec S_\chi\cdot\left(\frac{\vec q}{m_N}\times\vec v_\text{inel}^\perp\right),\\
&\mathcal{O}_6=\left(\vec S_\chi\cdot\frac{\vec q}{m_N}\right)\left(\vec S_N\cdot\frac{\vec q}{m_N}\right),\\[.2cm]
& \mathcal{O}_7=\vec S_N\cdot\vec v_\text{inel}^\perp,\quad \mathcal{O}_8=\vec S_\chi\cdot\vec v_\text{inel}^\perp,\\[.2cm]
&\mathcal{O}_9=i \vec S_\chi\cdot\left(\vec S_N\times\frac{\vec q}{m_N}\right),\quad \mathcal{O}_{10}=i \vec S_N\cdot\frac{\vec q}{m_N},\\[.2cm]
& \mathcal{O}_{11}=i \vec S_\chi\cdot\frac{\vec q}{m_N},\quad \mathcal{O}_{12} = \vec S_\chi\cdot\left(\vec S_N\times\vec v_\text{inel}^\perp\right), \\
&  \mathcal{O}_{13}= i \left(\vec S_\chi\cdot\vec v_\text{inel}^\perp\right)\left(\vec S_N\cdot\frac{\vec q}{m_N}\right),\\
& \mathcal{O}_{14} = i \left(\vec S_\chi\cdot\frac{\vec q}{m_N}\right)\left(\vec S_N\cdot\vec v_\text{inel}^\perp\right),\\
&  \mathcal{O}_{15} = -\left(\vec S_\chi\cdot\frac{\vec q}{m_N}\right)\left((\vec S_N\times\vec v_\text{inel}^\perp)\cdot\frac{\vec q}{m_N}\right),
\end{split}
\ee
where $\vec{S}_{\chi,N}$ are the spin operators for the dark matter and nucleon.  In \cite{Fitzpatrick:2012ix}, operator $\mathcal{O}_2$ was not considered since it doesn't appear in the nonrelativistic reduction of the scattering matrix elements of relativistic operators, and we find the same result here.  Thus, the important operators are at most linear in $\vperpinel$.  Since $\vperpinel$ differs from the elastic $\vec{v}^\perp$ by just a shift in $\vec{q}$, we will later find that this linearity allows one to utilize the form factors provided  by the Mathematica package \cite{Anand:2013yka}.

There are two other modifications to the elastic case that we will find.  First of all, $\delta$ can be a coefficient multiplying the operators when one reduces from relativistic operators.  The second effect is that $\vec{q}$ no longer has to appear in the combination of $i \vec{q}$, as can be seen by the expression for $\vperpinel$.  In the elastic case, this was guaranteed by the interaction being Hermitian.  Since conjugation swaps initial and final states, this acts as time reversal, $i \vec{q}\xrightarrow[]{T} i\vec{q}$.  However, for the inelastic case, the initial and final states are not the same particle, so this is no longer required by the interaction. In general, the inelastic operators in \eqnref{eqn:operators} may have arbitrary complex coefficients, as long as they appear in appropriate Hermitian conjugate pairs in the Hamiltonian. This was not the case for elastic operators because Hermiticity requires them to  have real coefficients.  

\subsection{Form Factors for Inelastic Scattering}
Now, one must use these nucleon-dark matter operators to determine the matrix elements within the target nucleus.  We will give a brief summary here, giving more details in Appendix \ref{app:Trans Amp}.  Since inelasticity modifies $\vperpinel$, we should examine how this affects the nuclear response.  First of all, by introducing the target velocity $\vec{v}_{T}$, we rewrite
\bea
\vec{v}^\perp_\text{el}& =& \vec{v}+\frac{\vec{q}}{2\mu_N} \\&=& \left(\frac{\vec{p}}{m_{\chi_1}} - \frac{\vec{k}}{m_N}\right) +\frac{1}{2m_{\chi_1}}\left(\vec{p}\,'-\vec{p}\,\right)+\frac{1}{2m_N}\left(\vec{k}-\vec{k}'\right) \nonumber \\
&\approx& \frac{1}{2}(\vec{v}_{\chi_1}+\vec{v}_{\chi_2}-\vec{v}_{N_{in}}-\vec{v}_{N_{out}}) \nonumber \\
& = & \frac{1}{2}(\vec{v}_{\chi_1}+\vec{v}_{\chi_2}-\vec{v}_{T_{in}}-\vec{v}_{T_{out}}) \nonumber \\&& +\frac{1}{2}\left[(\vec{v}_{T_{in}}-\vec{v}_{N_{in}})+(\vec{v}_{T_{out}}-\vec{v}_{N_{out}})\right]  \nonumber\\
&\equiv& \vec{v}_{\text{el}\, T}^\perp + \vec{v}_\text{nuc}.\nonumber
\label{eqn:NRops}
\eea
Thus for each nucleon in the nucleus, $\vec{v}^\perp_\text{el}$ is equal to the target's $\vec{v}^\perp_\text{el}$ plus a term, $\vec{v}_\text{nuc}$, that is dependent on the nucleon's relative velocity to the nucleus.  Similarly, for the inelastic velocity, we have
\bea
 \vperpinel &=& \vec{v}^\perp_{\text{inel}\, T} + \vec v_\text{nuc}
\eea 
where 
\bea
 \vec{v}^\perp_{\text{inel}\, T}  =  \frac{1}{2}(\vec{v}_{\chi_1}+\vec{v}_{\chi_2}-\vec{v}_{T_{in}}-\vec{v}_{T_{out}}) + \frac{ \delta}{|\vec{q}\,|^2}\vec{q},
\eea
Since the nucleus and dark matter scattering is also in the nonrelativistic limit, the same kinematic considerations from before show that $ \vec{v}^\perp_{\text{inel}\, T} $ is perpendicular to $\vec{q}$ and thus we can now interpret $\vec{q}$  as the momentum transfer from $\chi_1$ to the target {\em nucleus}.  

The reason for the separation of $\vec v_{\text{inel}\,T}^\perp$ into target and relative parts is that the nuclear form-factors only depend on interactions with {\em nucleons}, so only $\vec v_\text{nuc}$ is an operator.  The five nucleon interactions are \cite{Fitzpatrick:2012ix}:

\begin{equation}
\begin{split}
 \mathcal{O}_1^N&={\bf 1}_N,\quad\quad\mathcal{O}_2^N=-2\vec v_\text{nuc}\cdot\vec S_N,\\
 \mathcal{\vec O\text{$_3^N$}}&=2\vec S_N,\quad\quad\mathcal{\vec O\text{$_4^N$}}=-\vec v_\text{nuc}, \text{ and}\\
 \mathcal{\vec O\text{$_5^N$}}&=2i \vec v_\text{nuc}\times\vec S_N .
 \label{eq:Nuclear Ops}
 \end{split}
\end{equation}

\noindent which correspond to different types of nucleon responses.  $\mathcal{O}_1^N$ corresponds to the charge interaction, $\mathcal{O}_2^N$ to the axial charge interaction, $\mathcal{\vec O\text{$_3^N$}}$ to the axial vector interaction, $\mathcal{\vec O\text{$_4^N$}}$ to the vector magnetic interaction, and $\mathcal{\vec O\text{$_5^N$}}$ to the vector electric interaction.  Note that the explicit dependence on the inelastic nature of the scattering is not in the operators but in the coefficients.  For a more detailed discussion of the nuclear form factors see \cite{Fitzpatrick:2012ix}.

For our cases, since $\vperpinel$ only appears linearly (see Tables \ref{tbl:fermion}-\ref{tbl:scalarscalar}), we merely have to incorporate the change of $\vec{v}^\perp_{\text{el}\, T} \to \vec{v}^\perp_{\text{inel}\, T}$ in the Mathematica notebook \cite{Anand:2013yka}.  In calculating the matrix elements squared, this results in terms which are proportional to $|\vec{v}^\perp_{\text{inel}\, T}|^2$.  This has the simple form
\bea
|\vec{v}^\perp_{\text{inel}\, T}|^2 &=& |\vec{v}_T|^2-\left(\frac{1}{2\mu_T} + \frac{\delta}{|\vec{q}\,|^2}\right)^2 |\vec{q}\,|^2 \nonumber  \\
&=& |\vec{v}_T|^2 - v_{\text{min}\, T}^2
\label{eqn:vinelsq expanded}
\eea
where $\vec{v}_T = \vec{v}_{\chi_1} - \vec{v}_{T_{in}}$ and $\mu_T$ is the $\chi_1$-nucleus reduced mass.  In the second form, we have written the subtracted term as $v_{\text{min}\, T}$, the minimum speed to scatter off of the nucleus with energy $E_R$, which is the nucleus version of \eqnref{eqn:minv}.  Note that for upscattering ($\delta>0$) this leads to a suppression of this factor and for both signs of $\delta$, this term goes to zero at the minimum incoming velocity.   

The power of this formalism is that it gives the correct variables in which to characterize inelastic scattering and thus is helpful for understanding results that are at first surprising.  As an example, in Ref.~\cite{Chang:2010en}, an inelastic  dark matter model was analyzed that had a magnetic dipole interaction with the nucleus.  For the scattering of this dark matter dipole off of the nucleus charge,  peculiar terms involving $\delta/|\vec{v}|^2, \delta/E_R$ are found.  In that paper, these terms were only discovered by a systematic expansion.  However, in terms of this discussion, these terms are just due to the contribution from  the $\delta$ dependent terms of $|\vec{v}^\perp_{\text{inel}\, T}|^2$.   Of course, the main improvement on  Ref.~\cite{Chang:2010en} is that the form factors can now be reliably computed by a modification of the Mathematica notebook \cite{Anand:2013yka}.  Again, for details on how to implement these inelastic modifications to the form factor calculation, see Appendix \ref{app:Trans Amp}.

\subsection{Relativistic Matrix Elements for Fermion-Fermion Inelastic Transitions}
As a first application of this formalism, lets analyze the case where $\chi_{1,2}$ are both spin 1/2 fermions.  We start with the relativistic operators that would generate such scattering off of a nucleon.  We list the same twenty operators of \cite{Anand:2013yka} in Table \ref{tbl:fermion} for inelastic scattering \footnote{Our operator 20  has one less factor of $i$ due to a typo in \cite{Anand:2013yka}.}.  Factors of $i$ are set up so that if $\Psi_2 = \Psi_1$, the operator is Hermitian, thus allowing a convenient comparison to the elastic case by taking $\delta=0$.  The third column is the nonrelativistic limit of the matrix element after multiplying by $1/(4m_N m_\chi)$ to get to standard nonrelativistic normalization.  This matrix element is then decomposed in the final column in the basis of the fifteen nonrelativistic operators of \eqnref{eqn:NRops}. 

%%%%
\begin{table*}
\begin{footnotesize}
\centering
\renewcommand{\arraystretch}{1.6}
\begin{tabular}{| c | c | c | c |}\hline
  Index & Relativistic Operator & Nonrelativistic Limit $\times \frac{1}{4m_Nm_\chi}$  & $\sum_i c_i \mathcal{O}_i$ \\ \hline
  1  & $\bar\chi_2\chi_1\bar NN$ & ${\bf1}_\chi{\bf1}_N$ & $\mathcal{O}_1$\\ \hline
  
  2  & $i\bar\chi_2\chi_1\bar N\gamma^5N$ & $i\frac{\vec q}{m_N}\cdot\vec S_N$ & $\mathcal{O}_{10}$\\ \hline
  
  3  & $i\bar\chi_2\gamma^5\chi_1\bar NN$ & $-i\frac{\vec q}{m_\chi}\cdot\vec S_\chi$ & $-\frac{m_N}{m_\chi}\mathcal{O}_{11}$\\ \hline
  
  4  & $\bar\chi_2\gamma^5\chi_1\bar N\gamma^5N$ & $-(\frac{\vec q}{m_\chi}\cdot\vec S_\chi)(\frac{\vec q}{m_N}\cdot\vec S_N)$ & $-\frac{m_N}{m_\chi}\mathcal{O}_6$\\ \hline
  
  5  & $\bar\chi_2\gamma^\mu\chi_1\bar N\gamma_\mu N$ & ${\bf1}_\chi{\bf1}_N$ & $\mathcal{O}_1$ \\ \hline
  
  6  & $\bar\chi_2\gamma^\mu\chi_1\bar Ni\sigma_{\mu\nu}\frac{q^\nu}{m_M}N$ & $\begin{array}{c}\frac{|\qvec|^2}{2m_Nm_M}{\bf1}_\chi{\bf1}_N\\+2\left(\frac{\vec q}{m_\chi}\times\vec S_\chi+i\vec v_\text{inel}^\perp\right)\cdot\left(\frac{\vec q}{m_M}\times\vec S_N\right)\end{array}$ & $\begin{array}{c}\frac{|\qvec|^2}{2m_Nm_M}\left(\mathcal{O}_1+\frac{4m_N}{m_\chi}\mathcal{O}_4\right)\\-\frac{2m_N}{m_M}\left(\frac{m_N}{m_\chi}\mathcal{O}_6+\mathcal{O}_3\right)\end{array}$ \\ \hline
  
  7  & $\bar\chi_2\gamma^\mu\chi_1\bar N\gamma_\mu\gamma^5N$ & $\begin{array}{c}-2\vec S_N\cdot(\vec v_\text{inel}^\perp-\frac{\delta}{|\qvec|^2}\vec q)\\+2i\vec S_\chi\cdot(\vec S_N\times\frac{\vec q}{m_\chi})\end{array}$ & $-2\left(\mathcal{O}_7+i\frac{m_N\delta}{|\qvec|^2}\mathcal{O}_{10}-\frac{m_N}{m_\chi}\mathcal{O}_9\right)$ \\ \hline
  
  8  & $i\bar\chi_2\gamma^\mu\chi_1\bar Ni\sigma_{\mu\nu}\frac{q^\nu}{m_M}\gamma^5N$ & $2i\frac{\vec q}{m_M}\cdot\vec S_N$ & $\frac{2m_N}{m_M}\mathcal{O}_{10}$ \\ \hline
  
  9  & $\bar\chi_2i\sigma^{\mu\nu}\frac{q_\nu}{m_M}\chi_1\bar N\gamma_\mu N$ & $\begin{array}{c}-\frac{|\qvec|^2}{2m_\chi m_M}{\bf1}_\chi{\bf1}_N\\-2\left(\frac{\vec q}{m_N}\times\vec S_N+i\vec v_\text{inel}^\perp\right)\cdot\left(\frac{\vec q}{m_M}\times\vec S_\chi\right)\end{array}$ & $\begin{array}{c}-\frac{|\qvec|^2}{2m_\chi m_M}\left(\mathcal{O}_1+\frac{4m_\chi}{m_N}\mathcal{O}_4\right)\\+\frac{2m_N}{m_M}\left(\mathcal{O}_6+\mathcal{O}_5\right)\end{array}$ \\ \hline
  
  10 & $\bar\chi_2 i\sigma^{\mu\nu}\frac{q_\nu}{m_M}\chi_1\bar Ni\sigma_{\mu\alpha}\frac{q^\alpha}{m_M}N$ & $4\left(\frac{\vec q}{m_M}\times\vec S_\chi\right)\cdot\left(\frac{\vec q}{m_M}\times\vec S_N\right)$ & $\frac{4m_N^2}{m_M^2}\left(\frac{|\qvec|^2}{m_N^2}\mathcal{O}_4-\mathcal{O}_6\right)$ \\ \hline
  
  11 & $\bar\chi_2 i\sigma^{\mu\nu}\frac{q_\nu}{m_M}\chi_1\bar N\gamma_\mu\gamma^5N$ & $4i\left(\frac{\vec q}{m_M}\times\vec S_\chi\right)\cdot\vec S_N$ & $\frac{4m_N}{m_M}\mathcal{O}_9$ \\ \hline
  
  12 & $i\bar\chi_2 i\sigma^{\mu\nu}\frac{q_\nu}{m_M}\chi_1\bar Ni\sigma_{\mu\alpha}\frac{q^\alpha}{m_M}\gamma^5N$ & $-\left[i\frac{|\qvec|^2}{m_\chi m_M}-4\vec v_\text{inel}^\perp\cdot\left(\frac{\vec q}{m_M}\times\vec S_\chi\right)\right]\frac{\vec q}{m_M}\cdot\vec S_N$ & $\begin{array}{c}-\frac{|\qvec|^2}{m_M^2}\left(\frac{m_N}{m_\chi}\mathcal{O}_{10}+4\mathcal{O}_{12}\right)\\-\frac{4m_N^2}{m_M^2}\mathcal{O}_{15}\end{array}$ \\ \hline
  
  13 & $\bar\chi_2\gamma^\mu\gamma^5\chi_1\bar N\gamma_\mu N$ & $\begin{array}{c}2\left(\vec v_\text{inel}^\perp-\frac{\delta}{|\qvec|^2}\vec q\right)\cdot\vec S_\chi\\+2i\vec S_\chi\cdot\left(\vec S_N\times\frac{\vec q}{m_N}\right)\end{array}$ & $2\left(\mathcal{O}_8+\mathcal{O}_9+i\frac{m_N\delta}{|\qvec|^2}\mathcal{O}_{11}\right)$ \\ \hline
  
  14 & $\bar\chi_2\gamma^\mu\gamma^5\chi_1\bar Ni\sigma_{\mu\nu}\frac{q^\nu}{m_M}N$ & $4i\vec S_\chi\cdot\left(\frac{\vec q}{m_M}\times\vec S_N\right)$ & $-\frac{4m_N}{m_M}\mathcal{O}_9$\\ \hline
  
  15 & $\bar\chi_2\gamma^\mu\gamma^5\chi_1\bar N\gamma_\mu\gamma^5N$ & $-4\vec S_\chi\cdot\vec S_N$ & $-4\mathcal{O}_4$ \\ \hline
  
  16 & $i\bar\chi_2\gamma^\mu\gamma^5\chi_1\bar Ni\sigma_{\mu\nu}\frac{q^\nu}{m_M}\gamma^5N$ & $4i\frac{\vec q}{m_M}\cdot\vec S_N\left(\vec v_\text{inel}^\perp-\frac{\delta}{|\qvec|^2}\vec q\right)\cdot\vec S_\chi$ & $\frac{4m_N}{m_M}\left(\mathcal{O}_{13}-i\frac{m_N\delta}{|\qvec|^2}\mathcal{O}_6\right)$ \\ \hline
  
  17 & $i\bar\chi_2 i\sigma^{\mu\nu}\frac{q_\nu}{m_M}\gamma^5\chi_1\bar N\gamma_\mu N$ & $2i\frac{\vec q}{m_M}\cdot\vec S_\chi$ & $\frac{2m_N}{m_M}\mathcal{O}_{11}$ \\ \hline
  
  18 & $i\bar\chi_2 i\sigma^{\mu\nu}\frac{q_\nu}{m_M}\gamma^5\chi_1\bar Ni\sigma_{\mu\alpha}\frac{q^\alpha}{m_M}N$ & $\begin{array}{c}\vec S_\chi\cdot\frac{\vec q}{m_M}\left[i\frac{|\qvec|^2}{m_Nm_M}-4\vec v_\text{inel}^\perp\cdot(\frac{\vec q}{m_M}\times\vec S_N)\right]\\-4\frac{\delta}{m_M}\vec S_\chi\cdot(\frac{\vec q}{m_M}\times\vec S_N)\end{array}$ & $\begin{array}{c}\frac{m_N^2}{m_M^2}\left(\frac{|\qvec|^2}{m_N^2}\mathcal{O}_{11}+4\mathcal{O}_{15}\right)\\-i\frac{4m_N\delta}{m_M^2}\mathcal{O}_9\end{array}$ \\ \hline
  
  19 & $i\bar\chi_2 i\sigma^{\mu\nu}\frac{q_\nu}{m_M}\gamma^5\chi_1\bar N\gamma_\mu\gamma^5N$ & $\begin{array}{c}-4i\frac{\vec q}{m_M}\cdot\vec S_\chi\left(\vec v_\text{inel}^\perp-\frac{\delta}{|\qvec|^2}\vec q\right)\cdot\vec S_N\\-i\frac{4\delta}{m_M}\vec S_\chi\cdot\vec S_N\end{array}$ & $\begin{array}{c}-\frac{4m_N}{m_M}\left(\mathcal{O}_{14}-i\frac{m_N\delta}{|\qvec|^2}\mathcal{O}_6\right)\\-i\frac{4\delta}{m_M}\mathcal{O}_4\end{array}$ \\ \hline
  
  20 & $\bar\chi_2 i\sigma^{\mu\nu}\frac{q_\nu}{m_M}\gamma^5\chi_1\bar Ni\sigma_{\mu\alpha}\frac{q^\alpha}{m_M}\gamma^5N$ & $4\frac{\vec q}{m_M}\cdot\vec S_\chi\frac{\vec q}{m_M}\cdot\vec S_N$ & $\frac{4m_N^2}{m_M^2}\mathcal{O}_6$ \\ \hline
\end{tabular}
\renewcommand{\arraystretch}{1.0}
\end{footnotesize}
\caption{\label{tbl:fermion} Relativistic operators for inelastic transitions between two fermions $\chi_{1,2}$, their matrix element in the nonrelativistic limit multiplied by a factor of $1/(4m_\chi m_N)$, and their expansion in the basis of allowed scattering matrix-elements.}
\end{table*}
%%%

When calculating the matrix elements, we do not find explicit terms with $\vperpinel$, instead we get terms of $\vec{v}^\perp_\text{el}$.  This  is because the additional term of $\frac{\delta}{|\qvec|^2}\vec{q}$ has momentum factors in the denominator, which shouldn't appear in a matrix element of a contact operator.  However, many of these factors of $\vec{v}^\perp_\text{el}$ appear as $\vec{v}^\perp_\text{el} \cdot (\vec{q} \times \vec{S})$ which are equivalent to $\vec{v}^\perp_\text{inel} \cdot (\vec{q} \times \vec{S})$.  The other terms are of the form $\vec{v}^\perp_\text{el} \cdot \vec{S}$ which we rewrite as $(\vec{v}^\perp_\text{inel}-\frac{\delta}{|\qvec|^2}\vec{q}) \cdot \vec{S}.$  Writing the matrix elements in terms of $\vperpinel$ is convenient since it minimizes cross terms in the matrix element squared.  Note that in operators 18 and 19 there are additional terms proportional to  $\delta$ which are new nontrivial contributions to the scattering amplitude.  Amusingly, these contributions come from terms of $\frac{\delta}{|\qvec|^2}\vec{q}$ dotted into $\vec{q}$, canceling the $|\qvec|^2$ term in the denominator.  As a final check, we see that when we take $\delta=0$ we recover the elastic results  in \cite{Anand:2013yka}.      

\begin{figure}[htpb]
\begin{center}
\includegraphics[scale=0.6]{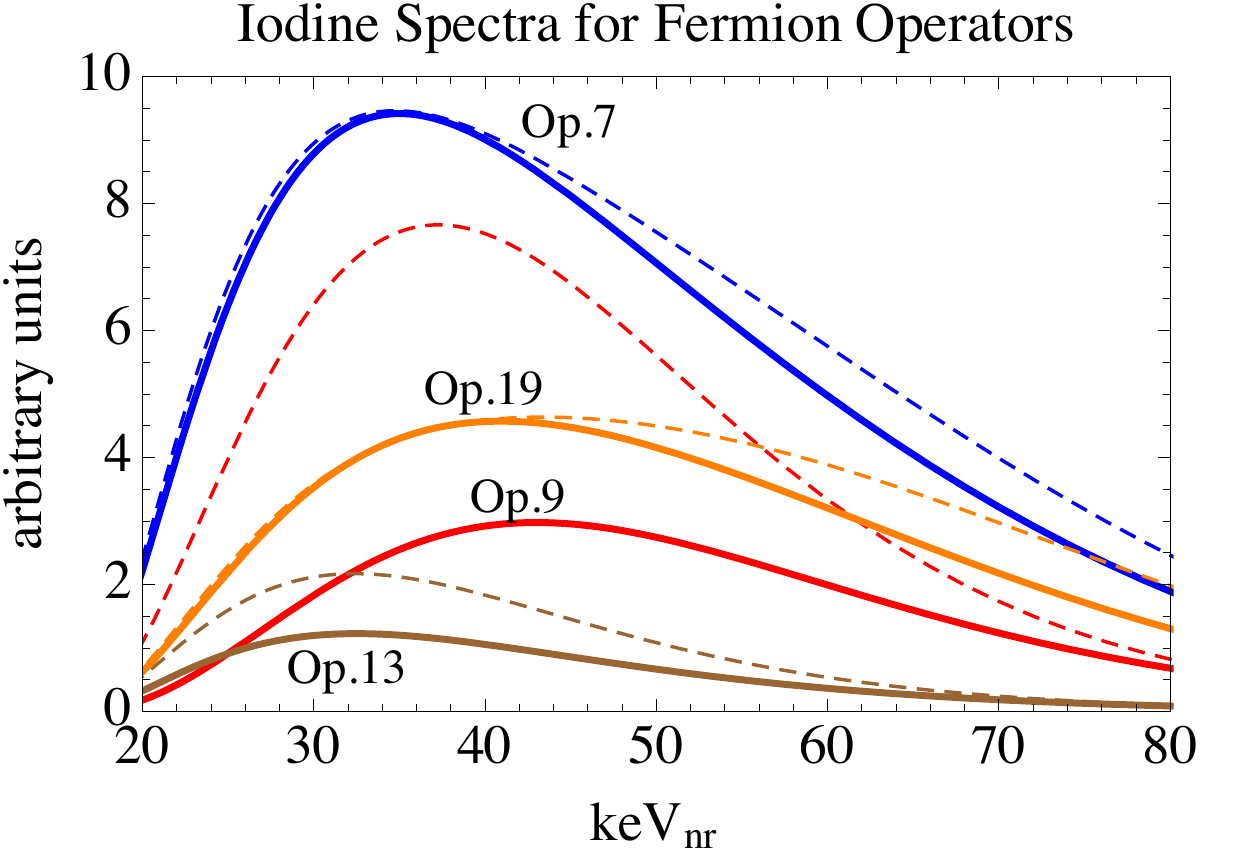}\hspace{.4in}
\end{center}
%\begin{minipage}{5.5in}
\caption{\label{fig:iodineexamples} Sample iodine scattering spectra with equal couplings to protons and neutrons for fermion operators $7, 9, 13, 19$.   The dark matter parameters are $m_\chi = 70$ GeV and $\delta = 120$ keV.  In solid are our predicted curves while dashed curves show incorrect spectra from combining elastic form factors with the inelastic velocity threshold.}
%\end{minipage}
\end{figure}

In \figref{fig:iodineexamples}, we plot some examples for the energy recoil spectra for these fermion operators in arbitrary units.  In this figure, we are assuming iodine scattering with equal couplings to protons and neutrons with a dark matter mass $m_\chi=70$ GeV and mass splitting $\delta=120$ keV.  In solid lines, we have our predicted rates.  As a comparison, we show in dashed lines an incorrect spectra if we had taken the elastic form factors but still integrated from the correct minimum velocity for inelastic scattering, $v_{\text{min}\, T}$.  Notice that the correct spectra is always smaller than the incorrect spectra which reflects the vanishing of $|\vec{v}^\perp_{\text{inel}\, T}|^2$ on threshold.  We chose these operators (7, 9, 13, 19) because they illustrate that the inelastic modifications to the form factors can in some cases significantly alter the shape and normalization of the spectra.  In addition, we found these differences to be quite sensitive to the choice of target nuclei and isospin structure of the nucleon couplings.

\subsection{Relativistic Matrix Elements for Scalar-Vector Inelastic Transitions}
%%%% Put in \chi!!!!!!!
\begin{table*}[htpb]
\begin{footnotesize}
\centering
\renewcommand\arraystretch{1.6}
\begin{tabular}{| c | c | c | c |}\hline
  Index & Relativistic Operator & Nonrelativistic Limit $\times\frac{1}{2m_N}$ & $\sum_i c_i \mathcal{O}_i$\\ \hline
  
  1  & $\frac{1}{m_M}\left(\Phi \overset\leftrightarrow{\partial}_\mu V^\mu\right)\bar NN$ & $i\frac{\vec q}{m_M}\cdot\vec\varepsilon$ & $\frac{m_N}{m_M}\mathcal{O}_{11}$ \\ \hline
  
  2  & $\frac{1}{m_M}\partial_\mu\!\left(\Phi V^\mu\right) \bar NN$ & $i\frac{\vec q}{m_M}\cdot\vec\varepsilon$ & $\frac{m_N}{m_M}\mathcal{O}_{11}$ \\ \hline
  
%  3  &   $\frac{K_\mu}{m_M}\Phi V^\mu\bar NN$ & $\frac{m_N}{m_M}\vec\varepsilon\cdot\left[2\vec v_\text{inel}^\perp-\left(\frac{2m_N\delta}{|\qvec|^2}-\frac{m_N}{m_\chi}\right)\frac{\vec q}{m_N}\right]$ & $\frac{m_N}{m_M}\left[2\mathcal{O}_8+i\left(\frac{2m_N\delta}{|\qvec|^2}-\frac{m_N}{m_\chi}\right)\mathcal{O}_{11}\right]$ \\ \hline
  3  &  $ \Phi V^\mu\bar N\gamma_\mu N$ & $\frac{1}{2} \vec\varepsilon\cdot\left[2\vec v_\text{inel}^\perp-\left(\frac{2m_N\delta}{|\qvec|^2}-\frac{m_N}{m_\chi}\right)\frac{\vec q}{m_N}\right]-i\vec\varepsilon\cdot(\frac{\vec q}{m_N}\times\vec S_N)$ & $  \mathcal{O}_8-\mathcal{O}_9+\frac{i}{2}\left(\frac{2m_N\delta}{|\qvec|^2}-\frac{m_N}{m_\chi}\right)\mathcal{O}_{11}$ \\ \hline
  4  & $\frac{i}{m_M}\left(\Phi \overset\leftrightarrow{\partial}_\mu V^\mu\right)\bar N\gamma^5N$ & $-\frac{\vec q}{m_M}\cdot\vec\varepsilon\frac{\vec q}{m_N}\cdot\vec S_N$ & $-\frac{m_N}{m_M}\mathcal{O}_6$ \\ \hline
  
  5  & $\frac{i}{m_M}\partial_\mu\!\left(\Phi V^\mu\right) \bar N\gamma^5N$ & $-\frac{\vec q}{m_M}\cdot\vec\varepsilon\frac{\vec q}{m_N}\cdot\vec S_N$ & $-\frac{m_N}{m_M}\mathcal{O}_6$ \\ \hline
  
  6 & $\Phi V^\mu\bar N\gamma_\mu\gamma^5N$ & $-2\vec S_N\cdot\vec\varepsilon$ & $-2\mathcal{O}_4$\\ \hline

  7  & $\Phi V^\mu\bar Ni\sigma_{\mu\nu}\frac{q^\nu}{m_M}N$ & $2i\vec\varepsilon\cdot(\frac{\vec q}{m_M}\times\vec S_N)$ & $-2\frac{m_N}{m_M}\mathcal{O}_9$ \\ \hline
  
  8  & $i\Phi V^\mu\bar Ni\sigma_{\mu\nu}\frac{q^\nu}{m_M}\gamma^5N$ & $i \vec\varepsilon\cdot\left[2\vec v_\text{inel}^\perp-\left(\frac{2m_N\delta}{|\qvec|^2}-\frac{m_N}{m_\chi}\right)\frac{\vec q}{m_N}\right]\frac{\vec q}{m_M}\cdot\vec S_N$ & $\frac{m_N}{m_M}\left[2\mathcal{O}_{13}-i\left(\frac{2m_N\delta}{|\qvec|^2}-\frac{m_N}{m_\chi}\right)\mathcal{O}_6\right]$ \\ \hline
  
\end{tabular}
\renewcommand{\arraystretch}{1.0}
\end{footnotesize}
\caption{\label{tbl:scalarvector} The inelastic relativistic operators for a transition from a dark matter particle of spin 0 to a spin 1 particle, $\Phi \to V^\mu$, their matrix element in the nonrelativistic limit after multiplying by a factor of $1/(2m_N)$, and then their decomposition in the basis of allowed scattering matrix elements.  This final step of replacing the spin 1 polarization vector $\vec{\epsilon}$ with $\vec{S}_\chi$, is valid if we multiply the final matrix element squared by a correction factor $c_{corr}$ in \eqnref{eqn:ccorr}.}
\end{table*}

An additional novelty of inelastic scattering is that it allows transitions between dark matter particles of different spin.  In this section, we consider the case where this transition is between a scalar $\Phi$ and a vector $V^\mu$.  Such nearly degenerate states have been shown to occur in models where the dark matter is composite \cite{Alves:2009nf, Kaplan:2009de} due to a hyperfine splitting in the dark sector.  In Table~\ref{tbl:scalarvector}, we list eight Hermitian operators which can be mediated by either spin 0 or 1 mediators.  For the third column, we list the matrix element's nonrelativistic limit after multiplying by a factor of $1/(2m_N)$ to go to the standard nonrelativistic normalization for the nucleons.    

All of these matrix elements are in the form of $M=  \vec{X}\cdot \vec{\epsilon}$, where $\vec{\epsilon}$ is the polarization vector of the spin 1 dark matter particle (which we take to be real for notational simplicity).  Depending on whether the spin 1 particle is in the initial or final state, we have to average or sum over these polarizations.  Since $\sum_{pol} \epsilon^{i\,} \epsilon^j = \delta^{ij}$, we have for the spin-summed (or averaged) matrix element squared
\bea
\overline{|M|^2}=   \Bigg\{ \renewcommand{\arraystretch}{1.3} \begin{array}{c} \frac{1}{3}  |\vec{X}|^2  \text{ spin 1 in initial state}\\  |\vec{X}|^2 \text{ spin 1 in final state}\end{array} .
\eea
This form allows us to treat these matrix elements with our basis of nonrelativistic operators in the following way.  If we just naively replace $\vec{\epsilon}$ with $\vec{S}_\chi$, we would have 
\bea
\overline{|M|^2}=\frac{1}{2s_\chi+1} \sum_{spins, i, j} S_\chi^i S_\chi^j X^{i*} X^j = \frac{s_\chi(s_\chi+1)}{3}|\vec{X}|^2. \nonumber \\[.2cm]
\eea
Thus, we can use the same operator basis where we naively replace $\vec{\epsilon}$ with $\vec{S}_\chi$ by multiplying the final result by a correction factor
\bea
\label{eqn:ccorr}
c_{corr} = \Bigg\{  \renewcommand{\arraystretch}{1.3} \begin{array}{c} \frac{1}{s_\chi(s_\chi+1)}  \text{ spin 1 in initial state}\\ \frac{3}{s_\chi(s_\chi+1)} \text{ spin 1 in final state}\end{array}.\nonumber \\[-.4cm]
\eea
Thus, in the final column of Table~\ref{tbl:scalarvector}, we decompose the matrix element under this replacement of $\vec{\epsilon}\to \vec{S}_\chi$, so that we can write it in the same operator basis as the fermion case.  These correction factors are accounted for in the additions we made to the Mathematica package of \cite{Anand:2013yka}.

\subsection{Relativistic Matrix Elements for Scalar-Scalar Inelastic Transitions}
As one more example, we analyze the case of a dark matter scattering process with a transition from a spin 0 particle $\Phi_1$ to another spin 0 particle $\Phi_2$.  In Table~\ref{tbl:scalarscalar}, we list seven operators between these two scalars which can be mediated by either spin 0 or 1 mediators.  For the third column, we list the matrix element's nonrelativistic limit after multiplying by a factor of $1/(2m_N)$ to go to the standard nonrelativistic normalization for the nucleons.  

\begin{table*}[htpb]
\begin{footnotesize}
\centering
\renewcommand{\arraystretch}{1.6}
\begin{tabular}{| c | c | c | c |}\hline
  Index & Relativistic Operator & Nonrelativistic Limit $\times\frac{1}{2m_N}$ & $\sum_i c_i \mathcal{O}_i$\\ \hline
  
  1  & $\Phi_2 \Phi_1 \bar NN$ & ${\bf1}_\chi{\bf1}_N$ & $\mathcal{O}_1$\\ \hline
  
  2  & $\Phi_2 \Phi_1 i \bar N\gamma^5 N$   & $i\frac{\vec q}{m_N}\cdot\vec S_N$ & $\mathcal{O}_{10}$\\ \hline
  
  3  & $\frac{1}{m_M} \left(i\Phi_2 \overset\leftrightarrow{\partial}_\mu \Phi_1\right) \bar N\gamma^\mu N$ & $ 2\frac{m_\chi}{m_M}{\bf1}_\chi{\bf1}_N$ & $2\frac{m_\chi}{m_M}\mathcal{O}_1$ \\ \hline
  
%  4  & $\frac{i}{m_M}  \partial_\mu\!\left(\Phi_2 \Phi_1\right) \bar N\gamma^\mu N$ & $-\frac{|\qvec|^4}{8m_M m_N^3} {\bf1}_\chi{\bf1}_N$ & $-\frac{|\qvec|^4}{8m_M m_N^3}\mathcal{O}_1$ \\ \hline
    
  4  & $\frac{1}{m_M} \left(i\Phi_2 \overset\leftrightarrow{\partial}_\mu \Phi_1\right) \bar N\gamma^\mu \gamma^5 N$ & $-4\frac{m_\chi}{m_M} \left(\vec v_\text{inel}^\perp-\frac{\delta}{|\qvec|^2}\vec q\right)\cdot\vec S_N$ & $-4\frac{m_\chi}{m_M}  \left(\mathcal{O}_7+i\frac{m_N\delta}{|\qvec|^2}\mathcal{O}_{10}\right) $ \\ \hline
  
  5  & $\frac{1}{m_M}\partial_\mu\!\left(\Phi_2 \Phi_1\right) \bar N\gamma^\mu \gamma^5 N$ & $- \frac{2i}{m_M} \vec q \cdot \vec S_N$ & $ -2 \frac{m_N}{m_M} \mathcal{O}_{10} $ \\ \hline
  
  6  & $\frac{1}{m_M} \left(i\Phi_2 \overset\leftrightarrow{\partial}_\mu \Phi_1\right) Ni\sigma_{\mu\nu}\frac{q^\nu}{m_M}N$ & $ 4i\frac{m_\chi}{m_M} \vperpinel \cdot \left(\frac{\vec q}{m_M} \times \vec{S}_N\right)+ \frac{m_\chi}{m_N m_M^2 }|\qvec|^2  {\bf1}_\chi{\bf1}_N$ & $  \frac{m_\chi}{m_N m_M^2 }|\qvec|^2 \mathcal{O}_1-4 \frac{m_\chi m_N}{m_M^2}\mathcal{O}_3$ \\ \hline
  
  7  & $\frac{i}{m_M} \left(i\Phi_2 \overset\leftrightarrow{\partial}_\mu \Phi_1\right) Ni\sigma_{\mu\nu}\gamma^5 \frac{q^\nu}{m_M}N$ & $4i\frac{m_\chi}{m_M} \frac{\vec q}{m_M} \cdot \vec S_N$ & $4\frac{m_\chi m_N}{m_M^2} \mathcal{O}_{10}$ \\ \hline
\end{tabular}
\renewcommand{\arraystretch}{1.0}
\end{footnotesize}
\caption{\label{tbl:scalarscalar}  The inelastic relativistic operators for a transition between dark matter particles both of spin 0, $\Phi_1 \to \Phi_2$, their matrix element in the nonrelativistic limit after multiplying by a factor of $1/(2m_N)$, and then their decomposition in the basis of allowed scattering matrix elements.}
\end{table*}

\section{Fitting DAMA/LIBRA's annual modulation signal \label{sec:dama}}

In this section we present fits to the DAMA/LIBRA annual modulation signal \cite{Bernabei:2013xsa} while also considering constraints from XENON10 \cite{Angle:2009xb},  XENON100 \cite{Aprile:2012nq}, LUX \cite{Akerib:2013tjd}, CDMS \cite{Ahmed:2010hw}, COUPP \cite{Behnke:2012ys}, and  KIMS \cite{Kim:2012rza}.   Unfortunately, we cannot be inclusive in our consideration of constraints. In particular we cannot derive limits from other direct detection experiments such as CRESST ($\text{CaWO}_4$) \cite{Angloher:2011uu} or fully analyze KIMS (CsI) which could be sensitive to the preferred parameter spaces.  This is because tungsten and cesium form factors are not yet available in the Mathematica package \cite{Anand:2013yka}, so we cannot treat them at the same level.  However, KIMS most recent analysis \cite{Kim:2012rza} claims any scenario involving iodine scattering to explain the DAMA modulation is incompatible with their data, which considering {\em only} iodine scattering, is mostly accurate, but there are some exceptions.  As we will demonstrate, KIMS limits are strongly dependent on the iodine quenching factors which have some large uncertainties at the moment.  Given all of these caveats, we will find some allowed regions on parameter space but expect these scenarios to be tested in the near future.

\subsection{Experimental Input}

To analyze the direct detection signal, we take a dark matter density $\rho = 0.3 \text{ GeV}/\text{cm}^3$ \cite{Bovy:2012tw} and a Maxwell-Boltzmann velocity distribution with parameters $v_0=220$ km/s \cite{1986MNRAS.221.1023K} and $v_\text{esc}=550$ km/s \cite{Smith:2006ym}.  For DAMA, since inelastic kinematics favors scattering off of heavier targets, we only consider scattering off of the iodine nuclei in the NaI crystals. We calculated the shift in the best fit points due to Na for operator 2 and found only a $0.07$\% change in the best fit $m_{M}$, and a $0.01$\% shift in $\chi^{2}$, so decided not to include Na in the full analysis. We found the modulation rate for scattering off of iodine alone and determined the point in $(m_\chi,\delta,m_M)$ parameter space which minimized a $\chi^{2}$ fit against the DAMA/LIBRA data \cite{Bernabei:2013xsa}. For our $\chi^2$, we used the first 12 bins of their data, which corresponds to an energy range of 2-8\keVee.  Later on, when we plot the 2D parameter space $(\delta,1/m_M)$, we will show contours for $\Delta \chi^2 =2.3, 5.99$ representing the $68, 95\%$ C.~L.\ region for two degrees of freedom ($d.o.f.$).

An important parameter in our fits is the quenching factor we adopt for iodine in NaI.  The quenching factor $Q$ determines the relationship between the measured energy in electron equivalents, keV$_\text{ee}$, and the original energy imparted to the nucleus keV$_\text{nr}$,  keV$_\text{ee} = Q\times\text{keV}_\text{nr}$.   Because of this, a good measurement of the quenching factor is necessary to determine the mass splitting and dark matter mass which best fits the DAMA/LIBRA modulation signal as well as  determining the constraints from other experiments.  For NaI, the value for iodine's quenching factor $Q_{I} = 0.09$ \cite{Bernabei:1996vj} is widely used, however a more recent paper \cite{Collar:2013gu} reports a measurement of $Q_{I} = 0.04$.  We will consider both values for iodine's quenching factor in what follows and denote it by $Q_\text{NaI}$.  A smaller quenching factor shifts the nuclear recoil energies that are relevant to DAMA to higher energies, so even though there is no suppression at xenon targets for scattering due to kinematics, the energy range could be outside of the acceptance range for LUX and XENON100 (this is more important for LUX as it has a smaller acceptance window).  We find that a smaller quenching factor generally requires a larger value of $\delta$ to fit the DAMA data which leads to a suppression of scattering at lighter targets like the germanium at CDMS.  These considerations mean that an uncertainty in the quenching factor has profound consequences for constraining signals seen in direct detection experiments. 

As limits, we first consider the xenon scattering limits in recent analyses by XENON100 \cite{Aprile:2012nq} and LUX \cite{Akerib:2013tjd}.   For XENON100's analysis, there was an exposure of $7.6\times 10^{3} \text{ kg}\cdot\text{days}$ and the acceptance we used was extracted from the hard discrimination cut of Fig. 1 in \cite{Aprile:2012nq} used in their maximum gap analysis.  This acceptance range is 2 to 43.3 keV$_\text{nr}$, though we extended their acceptance window to 50 keV$_\text{nr}$ assuming the acceptance didn't change in the last 6.7 keV$_\text{nr}$.  They observed two events, which we take to be all signal, giving a Poisson 90\% C.L. limit of 5.32 events.  LUX's analysis had $1.0\times 10^{4}\, \text{kg}\cdot\text{days}$ of exposure and used a 99.6\% efficiency after a 50\% NR acceptance in an energy range of 10-36 keV$_\text{nr}$ (the low energy, 0-10 keV$_\text{nr}$, efficiency isn't 99.6\% but can be found in the efficiency curve after the single scattering requirements have been accounted for in Fig. 1 of \cite{Akerib:2013tjd}).  They observed one event, which we take to be all signal, leading to a Poisson 90\% C.L. limit  of 3.89 events.  As both XENON100 and LUX experiments were primarily searching for elastic dark matter, their energy ranges weren't conducive to a search for inelastic dark matter which favors higher nuclear recoil energies, leading to weakened sensitivities.  To be sensitive to these high energy scatters, we also consider an older XENON10 analysis that was focused on inelastic dark matter \cite{Angle:2009xb}.  This XENON10 analysis had an exposure of $316 \text{ kg}\cdot\text{days}$, with an extended energy range of 75-250 keV$_\text{nr}$ that has a high efficiency $\sim 32\%$, after applying software cuts and nuclear recoil acceptance.  They saw no events in their extended range.  Since the advantage of this analysis over the more recent xenon experiments is its extended energy range and not its exposure we chose to constrain models if they predict more than 2.3 events (the 90\% C.L. limit with no observed events) in this 75-250 keV$_\text{nr}$ range. 

We looked at the constraints from CDMS inelastic dark matter search from their germanium detectors \cite{Ahmed:2010hw} as well.  Due to the lighter mass of germanium relative to xenon, we expected its limits would be suppressed relative to xenon limits.  This CDMS analysis had 970 \kgday\ of exposure, and even with perfect acceptance the exclusions for all operators were $\gtrsim1000$ times weaker than the limits from the xenon experiments.  Thus we decided not to include any more details for  germanium detectors.

An important constraint comes from COUPP which employs a CF$_3$I target \cite{Behnke:2012ys}.  We considered scattering of the dark matter off of the iodine as well as the fluorine, but not the carbon as its form factor isn't available in the Mathematica package.  However, due to carbon's light mass, it shouldn't give a significant contribution except for small mass splittings.    Our analysis of the COUPP data proceeds similarly to our analysis of the xenon experiments.  COUPP had three runs with i) exposures of 70.6 \kgday\ and an energy threshold of 7.8\keVnr, ii) 88.5 \kgday\ with an energy threshold of 11\keVnr, and iii) 394 \kgday\ with an energy threshold of 15.5\keVnr.  We considered only single bubble events for which there was a total efficiency of 79.1\%, and we used the step-function efficiency model \cite{Behnke:2013sma} for the iodine nucleation efficiency which rises to 100\% above 40 \keVnr.   Note that we didn't observe a significant shift in the derived limits when using the other parameterized efficiencies \cite{Behnke:2013sma}.  COUPP saw a total of 13 events for all three energy thresholds after time-isolation cuts.  Considering these as signal gives a Poisson 90\% C.L. limit of 18.96 events.  In all cases, we integrated scatters up to 200 keV$_\text{nr}$ which covers the range of allowed scatters.  

The last experiment we consider is KIMS \cite{Kim:2012rza} which has a CsI target.  Their analysis has $90\%$ C.L. limits on the dark matter scattering rate in eight bins ranging from 3-11 keV$_\text{ee}$.  For the purposes of constraining operators we consider a scenario ruled out if the predicted rate in any of these eight bins is larger than the stated limit for that bin.  Because KIMS uses CsI there is a different quenching factor for the iodine than the one for NaI crystals.  In \cite{Park:2002jr} the quenching factor is measured to be $\sim 0.10$ over a range of 20 to 120 keV$_\text{nr}$.  However, similar to  NaI, recent results \cite{Collar:2014lya} have pointed to a lower value of $Q_I\sim 0.05$ for CsI too.  The recent paper only measured CsI doped with sodium, which is not the same as the KIMS detectors which are doped with thallium.  However, in light of the new measurement and since the earlier measurement \cite{Park:2002jr} found similar quenching factors for detectors of different doping,  a value of $Q_I\sim 0.05$ for the KIMS detectors seems reasonable.  Thus, we consider both values in the following analysis and to differentiate it from the iodine quenching factor for NaI, we denote it as $Q_\text{CsI}$.  As another reminder, we emphasize that we cannot perform this analysis with cesium scattering, so all our constraints from the KIMS experiment are assuming only iodine recoils.   Thus, the KIMS limits should get stronger with cesium scattering, but we unfortunately  do not know how large of an effect this is.

One other issue we need to consider is the running time of these experiments, since large modulation can lead to order one changes in the scattering rate throughout the year.  We use the average scattering rate for XENON100, COUPP, and KIMS since their exposure was accumulated over a year, for LUX we use the maximum rate since its exposure was obtained during the summer, and for XENON10 we average over its run from October to February.  

\begin{table*}[htpb!]
\begin{footnotesize}
\centering
\renewcommand{\arraystretch}{1.6}
\begin{tabular}{c|ccc|c|rrrrrr}

 $Q_\text{NaI}$ & $m_{\chi} \mbox{(GeV)}$ & $\delta \mbox{(keV)}$ & $ m_{M} \mbox{(GeV)}$ & $\chi^{2}/d.o.f.$ & $r_\text{XENON10}^{90\%}$ & $r_\text{XENON100}^{90\%}$ & \quad $r_\text{LUX}^{90\%}$ & $r_\text{COUPP}^{90\%}$ & $r_\text{KIMS,0.10}^{90\%}$\ & $r_\text{KIMS,0.05}^{90\%}$\\ \hline 
\multicolumn{11}{c}{{Magnetic Inelastic Dark Matter}}\\%[1ex] \hline
\hline
 0.09 & 58.0 & 111.7 & 3209 & 0.97 & 0.0002 & 17.500 & 67.5 & 1.39 & 1.17 & 0.08 \\
  0.04 & 122.7 & 179.3 & 1096 & 0.82 & 0.2943 & 0.284 & 0.0 & 1.32 & 0.97 & 0.93 \\ \hline
\end{tabular}
\renewcommand{\arraystretch}{1.0}
\end{footnotesize}
\caption{This table presents the best fit parameters to the DAMA/LIBRA data for the magnetic dipole transition operator  and its $\chi^{2}/d.o.f.$ value.  For these  operators, the couplings are defined in \eqnref{eqn:midmcouplings}.  The final five columns give normalized limits, with the ratio of predicted to 90\% C.L. allowed counts for XENON10, XENON100, LUX, and COUPP and the largest ratio of the KIMS bins counts/\kgday/keV over the 90\% C.L. limit (this limit is presented for two different iodine quenching factors for KIMS, $Q_\text{CsI}$).  There is a fit for two values of the iodine quenching factor for NaI $Q_\text{NaI} = 0.09,\,0.04$.  Due to data taking conditions, the values for the XENON100, COUPP, and KIMS columns use the average yearly rate, the rate for LUX was the maximum, and the rate for XENON10 was averaged from October to February.}
\label{tbl:datamagnetic}
\end{table*}

\subsection{Reanalysis of Magnetic Inelastic Dark Matter}

In this section, we revisit the case of magnetic inelastic dark matter where the transition is mediated by a magnetic dipole transition \cite{Chang:2010en} 
\bea
{\cal L} = \frac{\mu_\chi}{2} \bar{\chi}_2 \sigma^{\mu\nu} \chi_1 F_{\mu\nu} + h.c.
\eea
Theoretically this scenario is appealing since the tensor operator vanishes for Majorana fermions, naturally leading to an inelastic transition.  Furthermore, iodine has a large dipole moment relative to most other heavy nuclear targets, mitigating xenon and tungsten constraints \cite{Chang:2010en}.  As mentioned earlier, the form factors used for these scenarios were highly uncertain \cite{Chang:2010en}, but we can now reliably calculate them with our modification of the Mathematica code.  Note that cesium does have a large dipole moment as well, but since it isn't implemented in the Mathematica notebook, we unfortunately have to neglect its scattering contribution.   

To calculate the form factor for the dipole transition, we use the following coefficients for the fermion operators 9 and 10 involving protons and neutrons
\begin{widetext}
\bea
 {\cal L}_\text{MIDM} = \frac{1}{q^2}\left[\bar\chi_2i\sigma^{\mu\nu}\frac{q_\nu}{m_M}\chi_1\; \bar p\gamma_\mu p\right]&+&0.9\frac{m_M}{m_N q^2}\left[\bar\chi_2 i\sigma^{\mu\nu}\frac{q_\nu}{m_M}\chi_1\; \bar p i\sigma_{\mu\alpha}\frac{q^\alpha}{m_M}p\right] \nonumber \\
 &-&0.96 \frac{m_M}{m_N q^2}\left[\bar\chi_2 i\sigma^{\mu\nu}\frac{q_\nu}{m_M}\chi_1\; \bar n i\sigma_{\mu\alpha}\frac{q^\alpha}{m_M}n\right]. \label{eqn:midmcouplings}
\eea
\end{widetext}
The relative coefficients are set by the proton and neutron magnetic moments being 2.8 and $-1.91$ nuclear magnetons, respectively.  Given the overall normalization, the relationship between our $m_M$ and the dark matter dipole moment is $1/m_M = e\mu_\chi$.

The best fit points in this parameter space are shown in Table \ref{tbl:datamagnetic} for the two choices of quenching factor, $Q_\text{NaI}=0.09,\,0.04$.  The $\chi^2/d.o.f.$ for our fit to DAMA is shown, with a $d.o.f.=9$, showing a very nice goodness of fit.  The final six columns show the normalized limits, $r$, from xenon and iodine experiments so that $r$ values above 1 are constrained at 90\% C.L. For XENON10, XENON100, LUX, and COUPP experiments, $r$ is the ratio of predicted events over the number of events allowed at 90\% C.L. (2.3, 5.32, 3.89, and 18.96 respectively).  For KIMS, in each bin from 3-11 keV$_\text{ee}$ we take the predicted bin rate divided by the 90\% C.L. limit on the rate in that bin, with $r$ being  the largest of these bin ratios.  We list KIMS constraints where we assume two values of the quenching factor  $Q_\text{CsI}=0.10$ and 0.05 for CsI.  Notice that for $Q_\text{NaI}=0.04$, the scenario is narrowly excluded by COUPP while being unconstrained by the other experiments.

\subsubsection*{Xenon Constraints}

The strength of the LUX or XENON100 limit depends strongly on the value of $Q_\text{NaI}$ we choose.  For the standard value $Q_\text{NaI} = 0.09$, the $2-6$\keVee\ energy range of DAMA's modulation spectra is $\sim22-67$\keVnr.  With the lower value of $Q_\text{NaI}= 0.04$ this changes to a much higher range of $50-150$\keVnr.  For inelastic dark matter, the modulated and unmodulated spectra span roughly the same energy bins and since xenon's mass is similar to iodine, the scattering off xenon will be roughly in the same range of nuclear recoil energies.  This explains why the LUX constraints are noticeably weaker for $Q_\text{NaI} = 0.04$, since its acceptance goes to zero above $\sim 36 \text{ keV}_\text{nr}$ while XENON100's goes up to 50 keV$_\text{nr}$.  This acceptance helps to make XENON100 competitive despite its smaller exposure.

\begin{figure*}[htb!]
\begin{footnotesize}
\centering
\renewcommand{\arraystretch}{1.6}
\begin{tabular}{cc}
\includegraphics[scale=0.81]{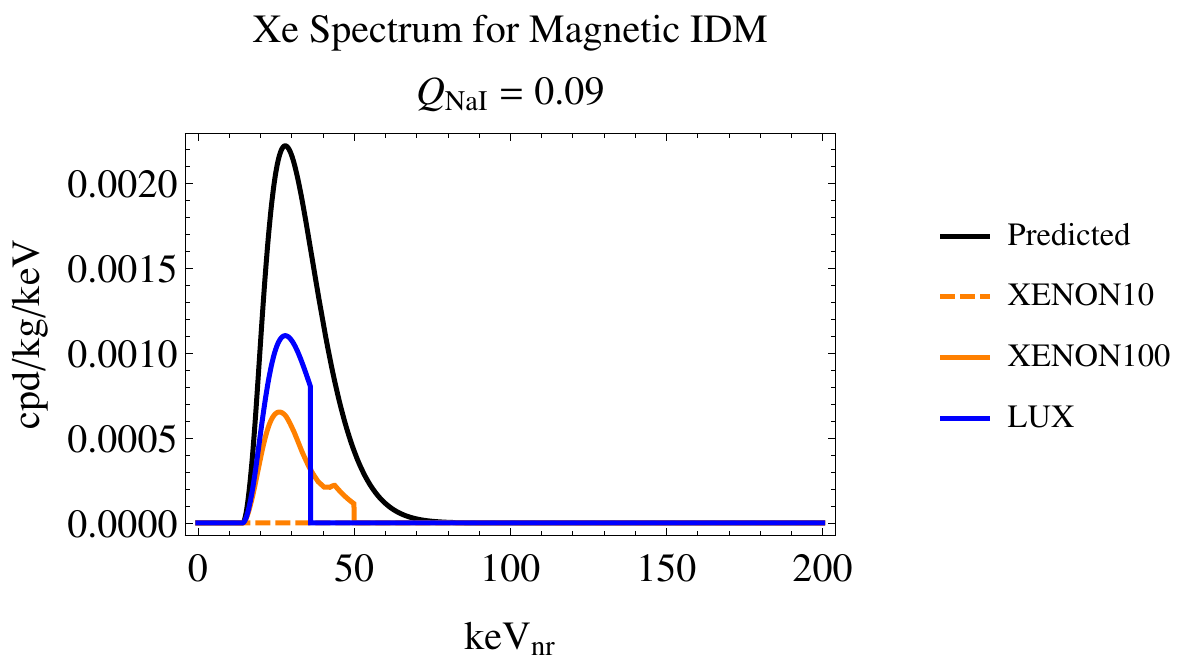}&\includegraphics[scale=0.61]{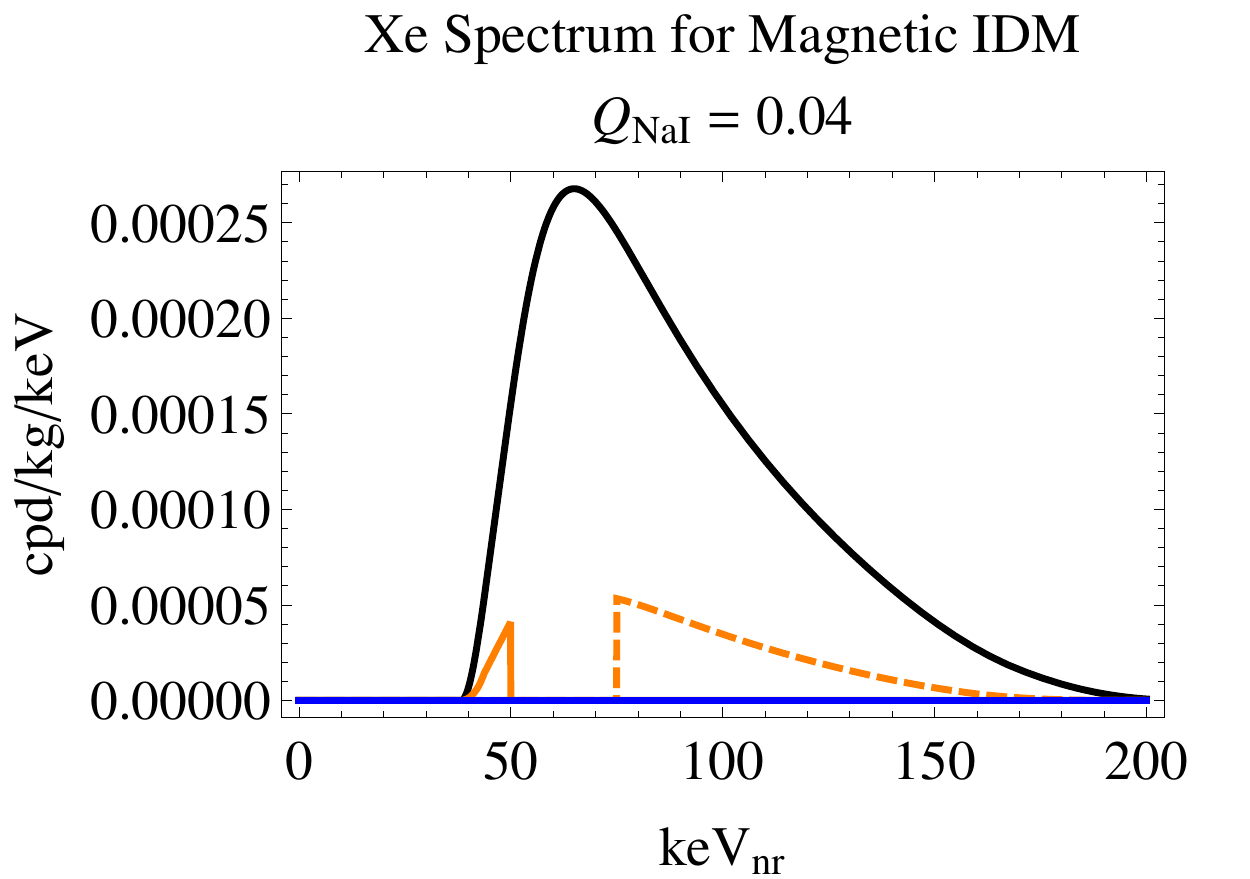}\\
\end{tabular}
\renewcommand{\arraystretch}{1.0}
\end{footnotesize}
\caption{These figures shows the xenon scattering spectrum for the best fit to DAMA's signal for magnetic inelastic dark matter  for two choices of $Q_\text{NaI}$.  The black curve is the expected spectrum while the orange (blue, orange-dashed) curve is the accepted spectrum for XENON100 (LUX, XENON10).  Note that for $Q_\text{NaI}=0.09$ the peak is visible to both XENON100 and LUX, but for $Q_\text{NaI}=0.04$ both these experiments' acceptances are too low at high energy to see a significant number of events.}
\label{fig:spectraXenon}
\end{figure*}

To show  this effect, we look at the best fit spectra for magnetic inelastic dark matter with different $Q_\text{NaI}$ values.  We saw that  XENON100 and LUX were a strong constraint for the larger value of the quenching factor, but the constraints for $Q_\text{NaI}=0.04$ were much weaker.  This is directly related to the location of the scattering spectrum relative to the experimental acceptance windows  as shown in \figref{fig:spectraXenon}.  For $Q_\text{NaI}=0.09$, the peak of the spectrum is well covered by both experiments, leading to the stringent constraints.   However, for $Q_\text{NaI}=0.04$, the peak scattering is missed by both experiments, with LUX having no sensitivity.   Given these high energy events, we also checked the constraints from XENON10's inelastic dark matter analysis \cite{Angle:2009xb} which extended to much higher energies.   In \figref{fig:spectraXenon} and Table~\ref{tbl:datamagnetic}, one can see that this XENON10 constraint is slightly stronger for the smaller iodine quenching factor, but is still not able to constrain this scenario due to its low exposure.   On the other hand, in existing XENON100 or LUX data there are about $\sim100$ events at high energy, so we encourage an extension of their analysis to energies above 50 keV$_\text{nr}$.  If the background in this region can be kept under control, they would have a high sensitivity to this scenario.

\subsubsection*{Iodine Constraints}

As expected, the constraints from other iodine detectors are very stringent for most inelastic dark matter scenarios since this is a direct comparison of the same target.  For COUPP constraints, changing $Q_\text{NaI}$ hardly affects the constraints.  The energy thresholds of the COUPP runs are not too high to lose many low energy events and the acceptance at high energy means that COUPP is sensitive to essentially all of the iodine scattering relevant for DAMA.  This explains why COUPP is the best constraint on DAMA both in terms of sensitivity and robustness from quenching factor uncertainties.       

\begin{figure*}[ht!]
\begin{footnotesize}
\centering
\renewcommand{\arraystretch}{1.6}
\begin{tabular}{cc}
\includegraphics[scale=0.55]{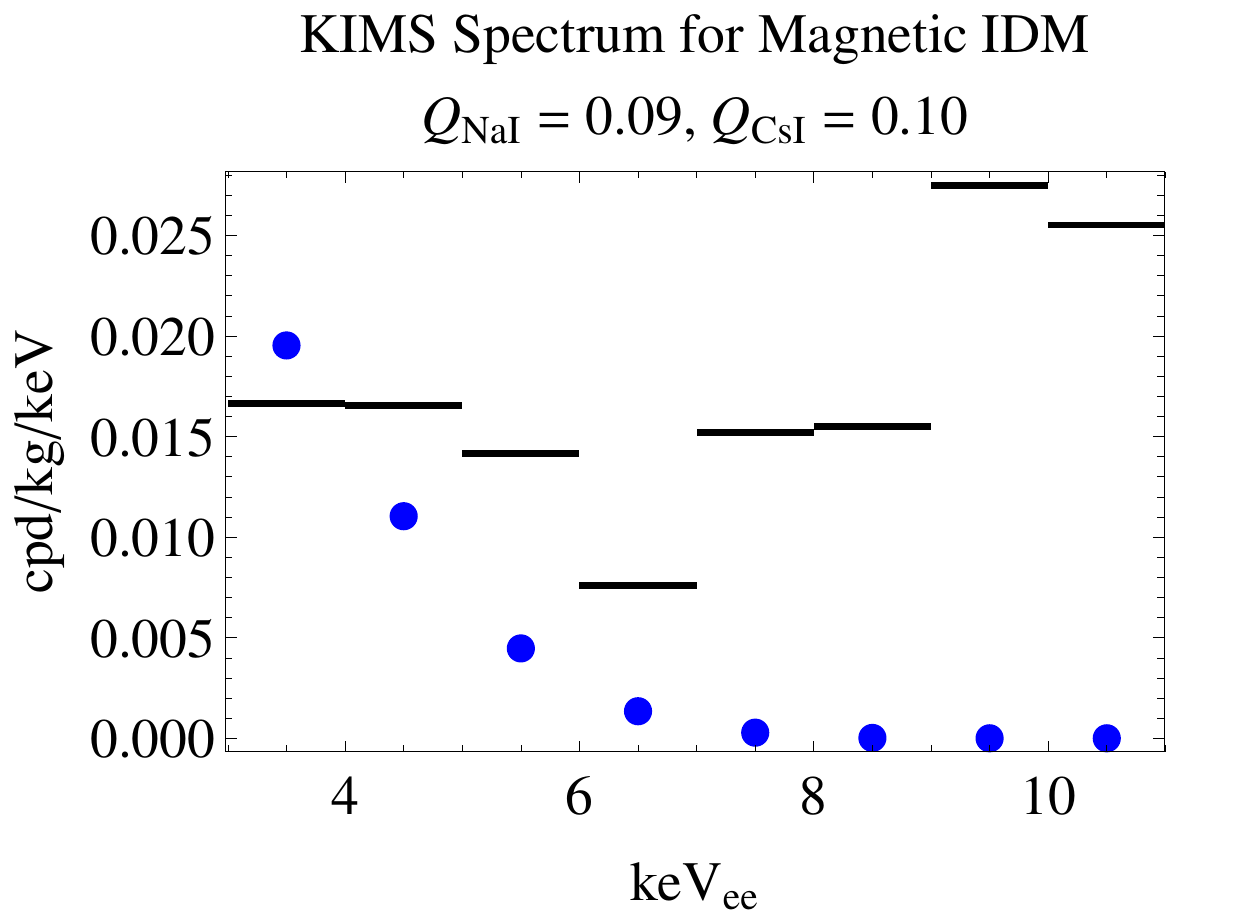}&\includegraphics[scale=0.55]{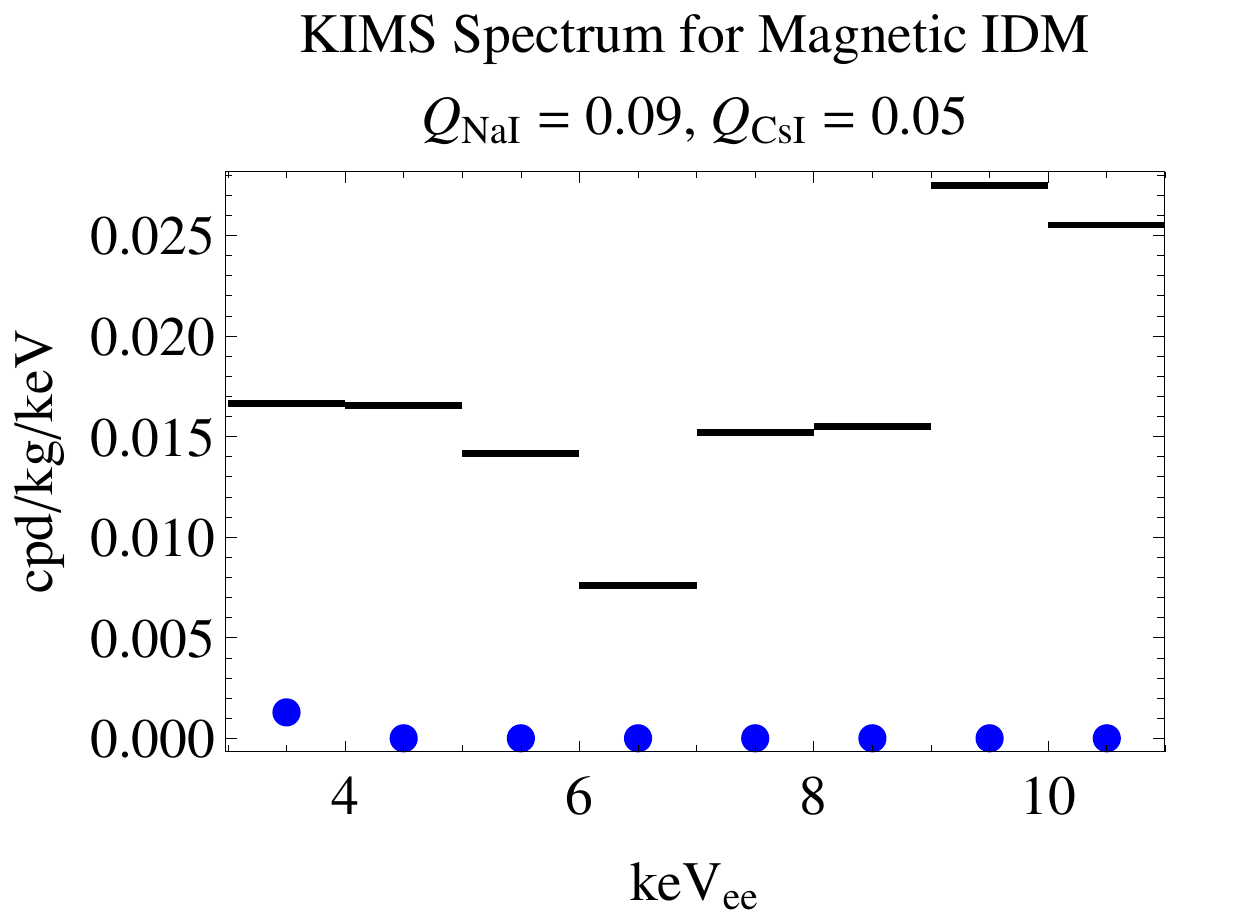}\\
\includegraphics[scale=0.55]{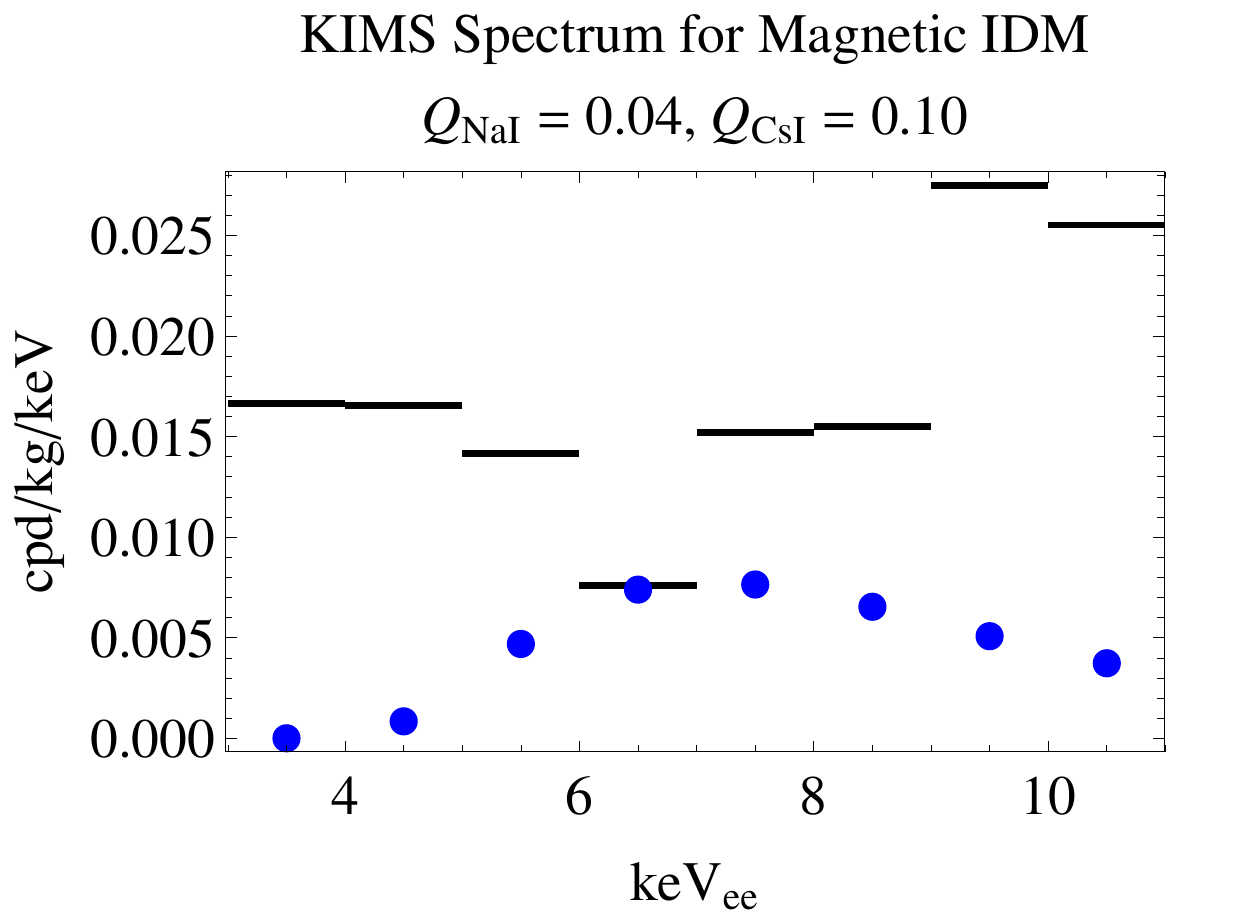}&\includegraphics[scale=0.55]{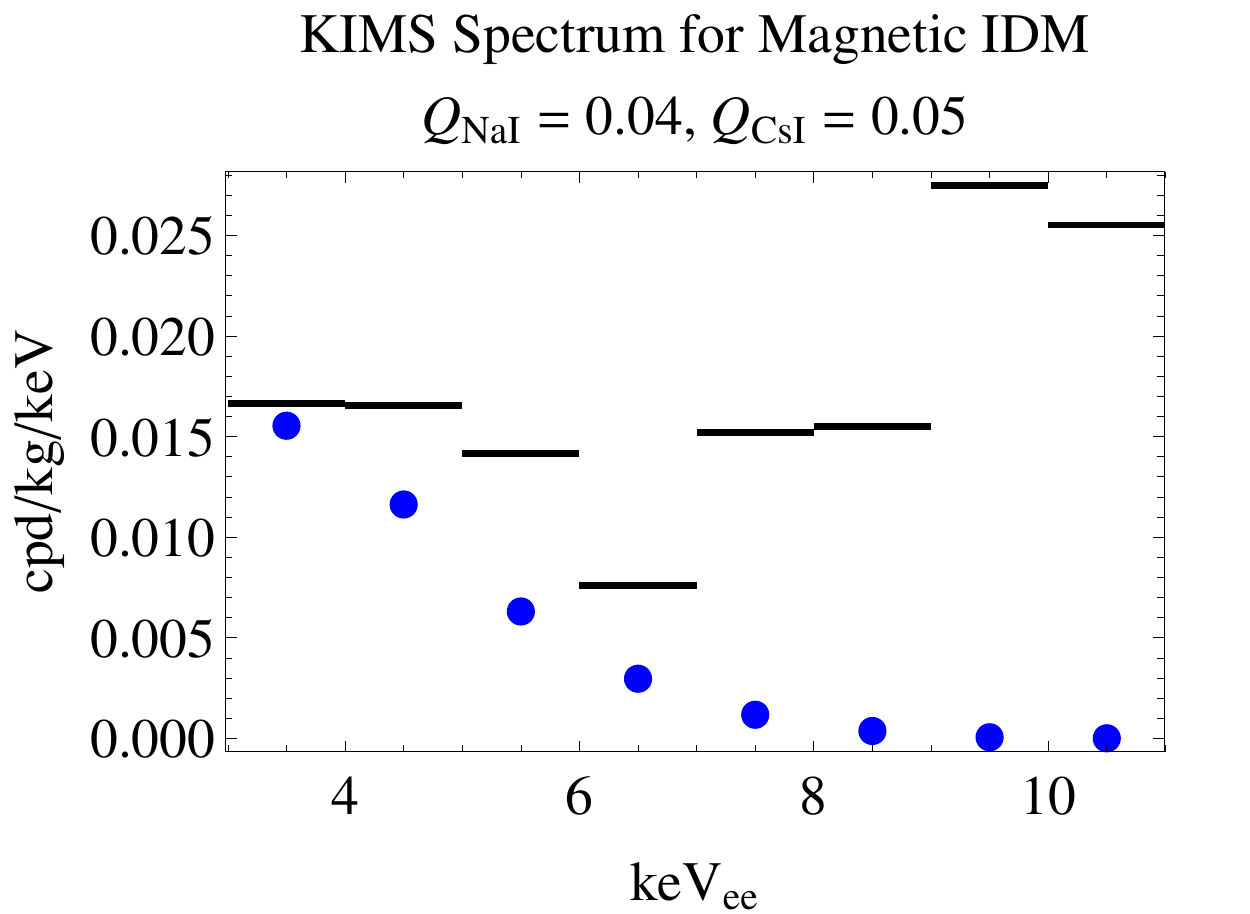}\\
\end{tabular}
\renewcommand{\arraystretch}{1.0}
\end{footnotesize}
\caption{This figure shows the KIMS energy spectrum for scattering events for magnetic inelastic dark matter at different $Q_\text{NaI}$'s and $Q_\text{CsI}$'s.  The blue points are the best fit points predicted rates and the black lines are the 90\% limits in each KIMS bin \cite{Kim:2012rza}.  Notice that the peak can shift from lower to higher energies as the quenching factors vary causing significant changes to the limit.}
\label{fig:spectraKIMS}
\end{figure*}

For KIMS, if the iodine quenching values used by the DAMA and KIMS experiments, $Q_\text{NaI}=0.09,Q_\text{CsI}=0.10 $ are correct, the  best fit point for magnetic inelastic dark matter is ruled out.  These constraints show a strong dependence on the quenching factor values chosen.  As the recent work of \cite{Collar:2013gu} and \cite{Collar:2014lya} shows, the correct values are not pinned down yet and could be significantly smaller.  This is especially relevant to KIMS constraints, since the scattering spectrum can be substantially shifted in energy, allowing much weaker constraints for some choices of the quenching factors. As an illustration,  we show in the four plots of \figref{fig:spectraKIMS} how the spectra at KIMS shifts  as we change the two quenching factors. In the upper left plot, we see that for the quenching factors $Q_\text{NaI} =0.09,Q_\text{CsI} =0.10$, the best fit point is constrained in the lowest KIMS bin.  However, in the upper right plot, changing to $Q_\text{CsI}=0.05$, we see that the spectrum shifts to energy bins below their threshold, giving no constraint.   In general, such a combination of quenching factors leads to particular weak limits from KIMS due to the scattering moving below threshold.   In the bottom left, the benchmark point with $Q_\text{NaI}=0.04,Q_\text{CsI}=0.10$, leads to a mild constraint in the 6 keV$_\text{ee}$ bin.   In the bottom right, changing the CsI quenching factor to 0.05, the spectrum shifts to lower values again leading to a rate that is almost constrained in the first bin with a smaller normalized limit, $r$.  Given the uncertainties, we consider both CsI quenching factors in presenting KIMS limits.  However,  if the same physics leads to the quenching factors of NaI and CsI to be of similar size, we find that KIMS becomes a more robust constraint.    

Up to these quenching factor issues, iodine targets still provide the  most model independent constraints on scenarios where iodine scattering explains the DAMA signal.  For these cases, the only way to suppress scattering is to have higher modulation amplitude.  Since COUPP and KIMS both ran over a year, this can lead to a modest drop in sensitivity which explains why the higher $\delta$ point has weaker constraints.    

\subsubsection*{Combined Limit Plots for Magnetic Inelastic Dark Matter}
\begin{figure*}[ht!]
 \begin{footnotesize}
 \centering
 \renewcommand{\arraystretch}{1.6}
 \begin{tabular}{cc}
\includegraphics[scale=0.55]{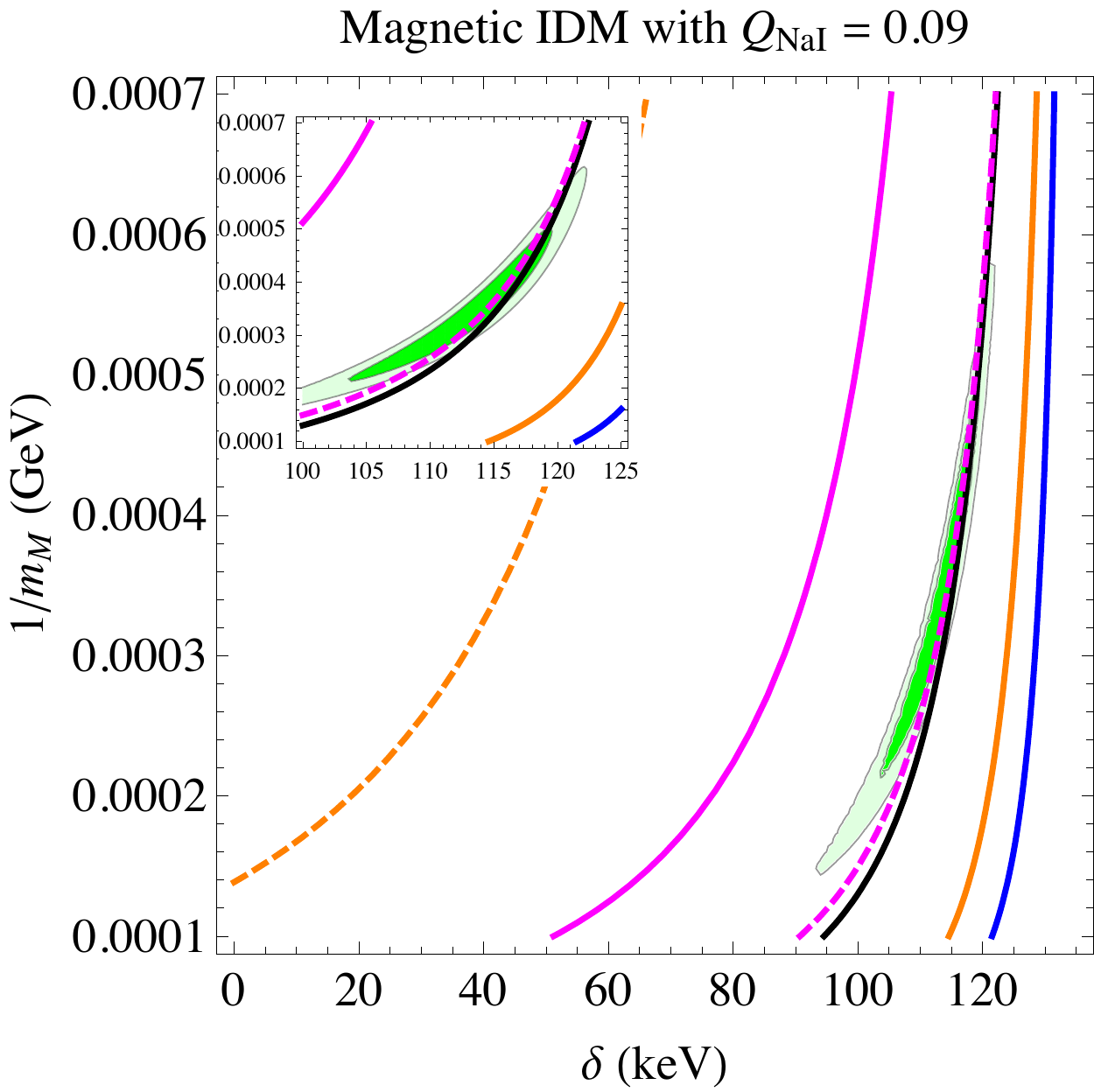}&\includegraphics[scale=0.56]{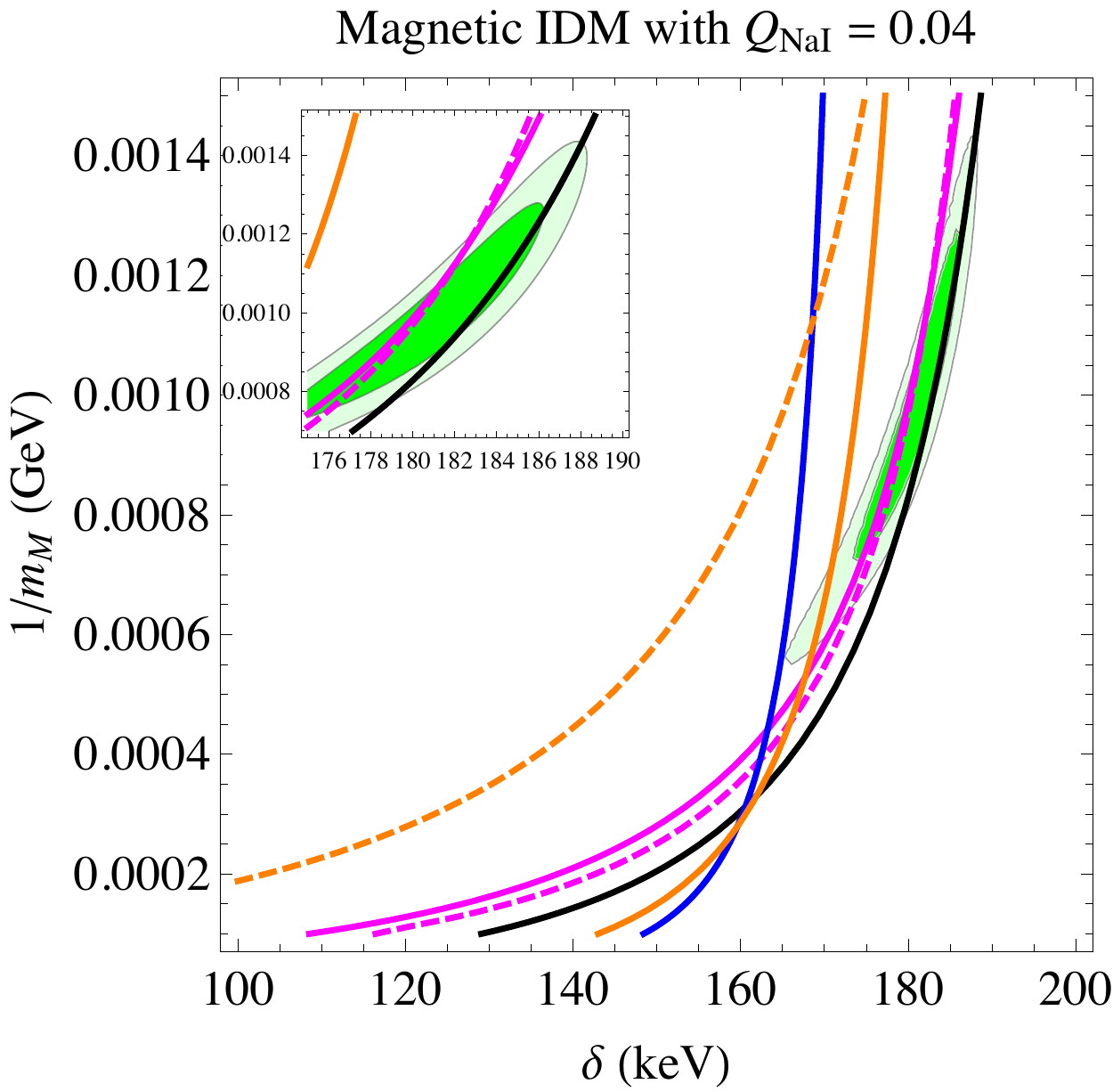}\\
 \end{tabular}
 \renewcommand{\arraystretch}{1.0}
 \end{footnotesize}
 \caption{This figure shows the combined limits plots for magnetic inelastic dark matter. The DM masses used are those listed with the corresponding operator in Table \ref{tbl:datamagnetic}. Constraints from LUX (blue), XENON100 (orange), XENON10 (orange dashed), KIMS ($Q_\text{CsI}=0.05$ magenta solid, $Q_\text{CsI}=0.10$ magenta dashed) and COUPP (black) are also shown, with the $90\%$ C.L. limits listed in section \ref{sec:dama}.}
\label{fig:combmidm}
 \end{figure*}

Although the best fit points for magnetic inelastic dark matter are ruled out conclusively by COUPP, there can be viable regions of parameter space which maintain a decent fit to DAMA.  To search for these we fix the best fit dark matter mass and then explored the remaining two dimensional parameter space in $(\delta,m_{M})$.  For DAMA, the $68, 95$\% C.L. parameter estimation regions were computed relative to the best fit $\chi^2$.  As can be seen in the left plot of \figref{fig:combmidm}, if $Q_\text{NaI}=0.09$, the constraints from LUX and XENON100 are strong and rule out all of the DAMA parameter space.  However, for the case of $Q_\text{NaI}=0.04$,  the right plot of \figref{fig:combmidm} shows that the constraints from all experiments weaken as one moves to higher values of the mass splitting, leading to a sliver of the $68$\% C.L. DAMA region which is not constrained and a significant region allowed at $95$\% C.L.  That XENON10 and the iodine experiments slowly fall off with increasing mass splitting shows how these experiments are mostly being weakened by increasing modulation and not a change in the energy spectrum.   

\begin{figure*}[ht!]
 \begin{footnotesize}
 \centering
 \renewcommand{\arraystretch}{1.6}
 \begin{tabular}{cc}
 \includegraphics[scale=0.55]{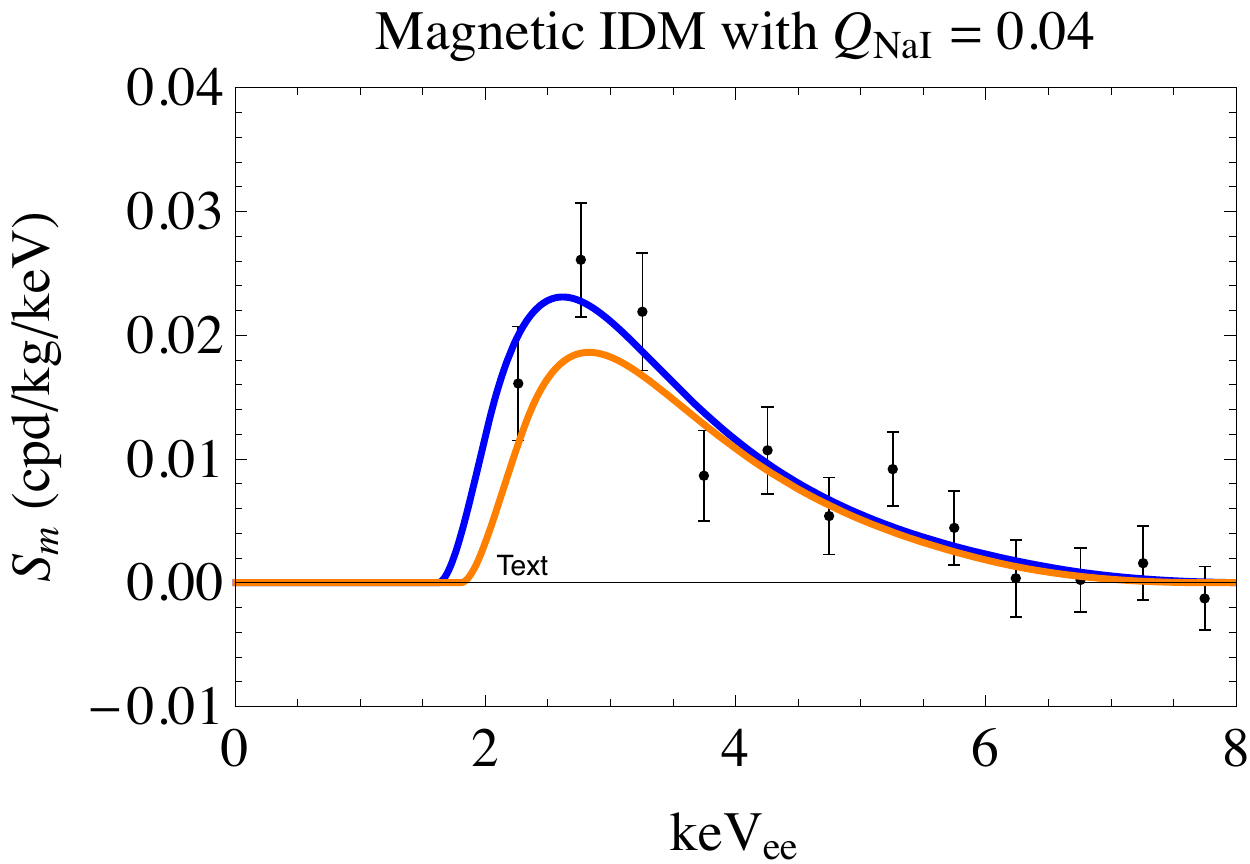}\\
 \end{tabular}
 \renewcommand{\arraystretch}{1.0}
 \end{footnotesize}
 \caption{This shows the magnetic inelastic dark matter  modulation amplitudes with the DAMA data points for comparison.  The plot assumes a iodine quenching factor $Q_\text{NaI} = 0.04$ and has both the best-fit modulation amplitude in blue and a sample unconstrained fit in orange.  The parameter values for the best fit are $(m_\chi,\delta,m_M) = (122.7\; \text{GeV}, 179.3\; \text{keV}, 1096 \; \text{GeV})$ and $\chi^{2}/d.o.f. = 0.82$   and  for the unconstrained point are $(m_\chi,\delta,m_M) = (122.7\; \text{GeV}, 184.5\; \text{keV}, 952 \; \text{GeV})$ and $\chi^{2}/d.o.f. = 1.17$.}
\label{fig:midmmod}
 \end{figure*}

In \figref{fig:midmmod}, we show the modulation spectra for the best fit point and an unconstrained point with the DAMA data points for comparison.  We see that the increase in mass splitting leads to a degradation in the $\chi^2$ but still has a good fit to the DAMA spectra.  Note that the values of $1/m_M$ required are quite reasonable since the magnetic moment of a particle should be of order a dark matter ``magneton" $=e/(2m_\chi)$, so that $1/m_M \sim e^2/(2m_\chi) = 5\times 10^{-4} (\frac{100 \GeV}{m_\chi}).$  The required magnetic moment seems to be similar to those seen in the nucleon sector and thus it seems plausible that this part of parameter space could appear generically in a complete model of magnetic inelastic dark matter.

\subsection{General Model Independent Analysis}
Now, we consider a more general model independent search for consistent scenarios that explain the DAMA annual modulation signal.   We performed a survey of the relativistic operators listed in Tables \ref{tbl:fermion}-\ref{tbl:scalarscalar} by analyzing the scattering when only one operator is turned on at a time.  Depending on the operator, we need to multiply by a dimensionful coupling $\lambda$ to describe the effective operator in the Lagrangian. For the fermion operators, we took this coupling to be $\lambda = 1/m_M^2$, so that $m_M$ characterizes the scale of the effective operator.  For the bosonic cases,  we instead take $\lambda = 1/m_M$.  Thus the parameters we varied were the dark matter mass $m_\chi$, the dimensional coupling parameter $m_M$, and the mass splitting  $\delta$. 

\begin{table*}[htpb]
\begin{footnotesize}
\centering
\renewcommand{\arraystretch}{1.6}
\begin{tabular}{cc|ccc|c|rrrrr}

Op. \# & $Q_\text{NaI}$ & $m_{\chi} \mbox{(GeV)}$ & $\delta \mbox{(keV)}$ & $ m_{M} \mbox{(GeV)}$ & $\chi^{2}/d.o.f.$ & $r_\text{XENON100}^{90\%}$ & \quad $r_\text{LUX}^{90\%}$ & $r_\text{COUPP}^{90\%}$ & $r_\text{KIMS,0.10}^{90\%}$\ & $r_\text{KIMS,0.05}^{90\%}$\\ \hline 
\multicolumn{11}{c}{{Spin $1/2 \to 1/2$ Transition}}\\%[1ex] \hline
\hline
 \multirow{2}{*}{2} & 0.09 & 44.2 & 59.0 & 15.9 & 1.08 & 0.049 & 0.130 & 3.28 & 2.27 & 0.21 \\
  & 0.04 & 84.6 & 103.2 & 17.4 & 1.03 & 0.006 & 0.014 & 4.23 & 2.50 & 2.16 \\ \hline
 
 \multirow{2}{*}{4} & 0.09 & 40.8 & 57.0 & 0.5 & 1.15 & 0.030 & 0.082 & 2.77 & 2.08 & 0.06 \\
  & 0.04 & 61.8 & 0.0 & 1.0 & 1.01 & 0.035 & 0.072 & 10.30 & 5.17 & 4.30 \\ \hline
 
 \multirow{2}{*}{7} & 0.09 & 55.3 & 108.3 & 1.6 & 0.97 & 0.017 & 0.053 & 1.41 & 1.09 & 0.11 \\
  & 0.04 & 111.8 & 163.2 & 1.5 & 0.97 & 0.001 & 0.000 & 1.81 & 1.22 & 1.12 \\ \hline
 
 \multirow{2}{*}{8} & 0.09 & 44.2 & 59.0 & 7.8 & 1.08 & 0.049 & 0.130 & 3.28 & 2.27 & 0.21 \\
  & 0.04 & 84.6 & 103.2 & 8.3 & 1.03 & 0.006 & 0.014 & 4.23 & 2.50 & 2.16 \\ \hline
 
 \multirow{2}{*}{9} & 0.09 & 47.4 & 69.0 & 4.6 & 1.01 & 40.420 & 117.900 & 3.00 & 2.19 & 0.21 \\
  & 0.04 & 95.1 & 135.1 & 4.0 & 0.99 & 1.791 & 2.847 & 2.79 & 1.83 & 1.59 \\ \hline
 
 \multirow{2}{*}{10} & 0.09 & 40.9 & 53.2 & 2.5 & 1.06 & 0.111 & 0.284 & 3.12 & 2.21 & 0.09 \\
  & 0.04 & 62.4 & 0.0 & 3.7 & 1.00 & 0.156 & 0.282 & 12.67 & 5.26 & 4.34 \\ \hline
 
 \multirow{2}{*}{11} & 0.09 & 50.8 & 96.6 & 5.8 & 1.11 & 0.042 & 0.119 & 1.72 & 1.29 & 0.10 \\
  & 0.04 & 85.5 & 106.4 & 7.7 & 1.03 & 0.032 & 0.056 & 4.26 & 2.44 & 2.10 \\ \hline
 
 \multirow{2}{*}{13} & 0.09 & 56.1 & 110.9 & 19.6 & 0.96 & 13.640 & 55.640 & 1.32 & 1.08 & 0.08 \\
  & 0.04 & 112.6 & 170.7 & 15.6 & 0.94 & 0.751 & 0.015 & 1.42 & 0.97 & 0.92 \\ \hline
 
 \multirow{2}{*}{14} & 0.09 & 50.8 & 96.6 & 5.8 & 1.11 & 0.042 & 0.119 & 1.72 & 1.29 & 0.10 \\
  & 0.04 & 85.5 & 106.4 & 7.7 & 1.03 & 0.032 & 0.056 & 4.26 & 2.44 & 2.10 \\ \hline
 
 \multirow{2}{*}{15} & 0.09 & 54.3 & 106.0 & 49.6 & 1.02 & 0.021 & 0.069 & 1.49 & 1.15 & 0.10 \\
  & 0.04 & 102.7 & 146.8 & 54.3 & 1.00 & 0.005 & 0.006 & 2.47 & 1.60 & 1.44 \\ \hline
 
 \multirow{2}{*}{19} & 0.09 & 52.6 & 97.5 & 0.6 & 1.02 & 0.045 & 0.139 & 1.66 & 1.23 & 0.09 \\
  & 0.04 & 99.9 & 137.7 & 0.7 & 1.01 & 0.012 & 0.015 & 2.18 & 1.36 & 1.20 \\ \hline
 
 \multirow{2}{*}{20} & 0.09 & 40.8 & 57.0 & 2.5 & 1.15 & 0.030 & 0.082 & 2.77 & 2.08 & 0.06 \\
  & 0.04 & 61.8 & 0.0 & 3.9 & 1.01 & 0.035 & 0.072 & 10.30 & 5.17 & 4.30 \\ \hline
\end{tabular}
\renewcommand{\arraystretch}{1.0}
\end{footnotesize}
\caption{This table presents the best fit parameters to the DAMA/LIBRA data for each fermionic operator for which iodine showed an enhancement over xenon, considering proton coupling {\em only},  and their $\chi^{2}/d.o.f.$ value.  For these fermion operators, the coupling is $\lambda = 1/m_M^2$.  The final five columns give normalized limits, with the ratio of predicted to 90\% C.L. allowed counts for XENON100, LUX, and COUPP and the largest ratio of the KIMS bins counts/\kgday/keV over the 90\% C.L. limit (this limit is presented for two different iodine quenching factors for KIMS, $Q_\text{CsI}$).  Each operator has a fit for two values of the iodine quenching factor for NaI $Q_\text{NaI} = 0.09,\,0.04$.  Due to data taking conditions, the values for the XENON100, COUPP, and KIMS columns uses the average yearly rate, and the rate for LUX was the maximum.}
\label{tbl:datafermion}
\end{table*}

To narrow our survey and to specifically avoid the stringent constraints of xenon target experiments, we only considered operators whose transition probabilities for iodine were significantly ($\geq10$ times) enhanced over xenon.  These operators were identified by examining the ratio of iodine's transition probability to xenon's at the minimum velocity for iodine (see \eqnref{eqn:minv}), as it is higher than the minimum velocity for xenon scattering.  This ratio was plotted, for a specific value of $m_\chi$ with the coupling $m_M$ set to 1, on the $(\delta,E_R)$ plane with $E_R$ the nuclear recoil energy.  The operators' coupling to nucleons was varied between pure proton, pure neutron, equal coupling to proton and neutron, and equal but opposite couplings.  We found that only pure coupling to protons significantly favored iodine over xenon and further that all iodine-enhanced operators had some contribution from the nucleon spin $\mathcal{\vec O\text{$_3^N$}}$, see \eqnref{eq:Nuclear Ops}.  Since iodine's nucleus has an unpaired proton while xenon has an unpaired neutron, this explains why the sensitivity is enhanced if we only couple to the proton \cite{Kopp:2009qt}.  As a check that this method for selecting operators finds all relevant ones, we also performed a full analysis for several other operators and nucleon couplings and found the results matched our predictions from this selection process. Note that our inability to treat cesium in KIMS is particularly important for coupling to proton spin, since cesium also has an unpaired proton.  On the other hand, tungsten isotopes only have unpaired neutrons, so we expect that their rates would be suppressed much like xenon targets. 

The best fit points in this parameter space is shown in Tables \ref{tbl:datafermion} and \ref{tbl:databoson} for the two choices of quenching factor of $Q_\text{NaI}=0.09,\,0.04$.  The $\chi^2/d.o.f.$ for our fit to DAMA is shown, with a $d.o.f.=9$, showing a reasonable goodness of fit for all operators.  The final five columns show the normalized limits, $r$, from xenon and iodine experiments so that $r$ values above 1 are constrained at 90\% C.L. For XENON100, LUX, and COUPP experiments, $r$ is the ratio of predicted events over the number of events allowed at 90\% C.L. (5.32, 3.89, and 18.96 respectively).  For KIMS, in each bin from 3-11 keV$_\text{ee}$ we take the predicted bin rate divided by the 90\% C.L. limit on the rate in that bin, with $r$ being  the largest of these bin ratios.  We list KIMS constraints where we assume two values of the quenching factor  $Q_\text{CsI}=0.10$ and 0.05 for CsI.  Notice that there are a few operators which are narrowly excluded by COUPP while being unconstrained by the other experiments.

Even though we've discussed how XENON10 is sensitive to much higher energy scatters than XENON100 or LUX, we find that it generically sets weaker constraints for this model independent analysis due to its lower exposure.  In a few cases, the limits of XENON10 were similar or just a bit larger than XENON100, for example fermion operators 7, 15, and 19, spin 0 to 1 operators 6, and spin 0 to 0 operator 4,  but they were not large enough to be constraining.  Because these constraints were not strong enough to rule out any best fit points, we chose not to include the XENON10 limits in our tables or figures for this model independent survey.

\begin{table*}[htpb]
\begin{footnotesize}
\centering
\renewcommand{\arraystretch}{1.6}
\begin{tabular}{cc|ccc|c|rrrrr}

Op. \# & $Q_\text{NaI}$ & $m_{\chi} \mbox{(GeV)}$ & $\delta \mbox{(keV)}$ & $ m_{M} \mbox{(GeV)}$ & $\chi^{2}/d.o.f.$ & $r_\text{XENON100}^{90\%}$ &\quad $r_\text{LUX}^{90\%}$ & $r_\text{COUPP}^{90\%}$ & $r_\text{KIMS,0.10}^{90\%}$\ & $r_\text{KIMS,0.05}^{90\%}$\\ \hline 
\multicolumn{11}{c}{{Spin $0 \to 1$ Transition}}\\%[1ex] \hline
\hline
 \multirow{2}{*}{4} & 0.09 & 40.8 & 57.0 & 0.5 & 1.15 & 0.030 & 0.082 & 2.77 & 2.08 & 0.06 \\
  & 0.04 & 61.8 & 0.0 & 1.0 & 1.01 & 0.035 & 0.072 & 10.30 & 5.17 & 4.30 \\ \hline
 
 \multirow{2}{*}{5} & 0.09 & 40.8 & 57.0 & 0.5 & 1.15 & 0.030 & 0.082 & 2.77 & 2.08 & 0.06 \\
  & 0.04 & 61.8 & 0.0 & 1.0 & 1.01 & 0.035 & 0.072 & 10.30 & 5.17 & 4.30 \\ \hline
 
 \multirow{2}{*}{6} & 0.09 & 54.3 & 106.0 & 22.6 & 1.02 & 0.021 & 0.069 & 1.49 & 1.15 & 0.10 \\
  & 0.04 & 102.7 & 146.8 & 14.3 & 1.00 & 0.005 & 0.006 & 2.47 & 1.60 & 1.44 \\ \hline
 
 \multirow{2}{*}{7} & 0.09 & 50.8 & 96.6 & 1.4 & 1.11 & 0.042 & 0.119 & 1.72 & 1.29 & 0.10 \\
  & 0.04 & 85.5 & 106.4 & 1.6 & 1.03 & 0.032 & 0.056 & 4.26 & 2.44 & 2.10 \\ \hline
%\multicolumn{7}{|c|}{}\\[-2.9ex]
\multicolumn{11}{c}{{Spin $0 \to 0$ Transition}}\\%[1ex] \hline
\hline 
 \multirow{2}{*}{2} & 0.09 & 44.2 & 58.2 & 1.0 & 1.08 & 0.049 & 0.127 & 3.32 & 2.29 & 0.21 \\
  & 0.04 & 84.6 & 103.3 & 1.1 & 1.03 & 0.005 & 0.013 & 4.21 & 2.49 & 2.15 \\ \hline
 
 \multirow{2}{*}{4} & 0.09 & 56.6 & 108.7 & 1.0 & 0.99 & 0.011 & 0.041 & 1.48 & 1.16 & 0.10 \\
 & 0.04 & 115.6 & 166.7 & 1.3 & 0.96 & 0.001 & 0.000 & 1.76 & 1.21 & 1.12 \\ \hline
 
 \multirow{2}{*}{5} & 0.09 & 44.2 & 58.2 & 1.0 & 1.08 & 0.049 & 0.127 & 3.32 & 2.29 & 0.21 \\
  & 0.04 & 84.6 & 103.3 & 1.1 & 1.03 & 0.005 & 0.013 & 4.21 & 2.49 & 2.15 \\ \hline
 
 \multirow{2}{*}{7} & 0.09 & 44.2 & 58.2 & 3.7 & 1.08 & 0.049 & 0.127 & 3.32 & 2.29 & 0.21 \\
  & 0.04 & 84.6 & 103.3 & 4.5 & 1.03 & 0.005 & 0.013 & 4.21 & 2.49 & 2.15 \\ \hline
\end{tabular}
\renewcommand{\arraystretch}{1.0}
\end{footnotesize}
\caption{This table presents the best fit parameters to the DAMA/LIBRA data for each bosonic operator for which iodine showed an enhancement over xenon, considering proton coupling {\em only},  and their $\chi^{2}/d.o.f.$ value.  For these bosonic operators, the coupling is $\lambda = 1/m_M$.  The final five columns give normalized limits, with the ratio of predicted to 90\% C.L. allowed counts for XENON100, LUX, and COUPP and the largest ratio of the KIMS bins counts/\kgday/keV over the 90\% C.L. limit (this limit is presented for two different iodine quenching factors for KIMS, $Q_\text{CsI}$).  Each operator has a fit for two values of the iodine quenching factor for NaI $Q_\text{NaI} = 0.09,\,0.04$.  Due to data taking conditions, the values for the XENON100, COUPP, and KIMS columns uses the average yearly rate, and the rate for LUX was the maximum.}
\label{tbl:databoson}
\end{table*}

\subsubsection*{Combined Limit Plots for Relativistic Operators}
\begin{figure*}[ht!]
 \begin{footnotesize}
 \centering
 \renewcommand{\arraystretch}{1.6}
 \begin{tabular}{cc}
\includegraphics[scale=0.55]{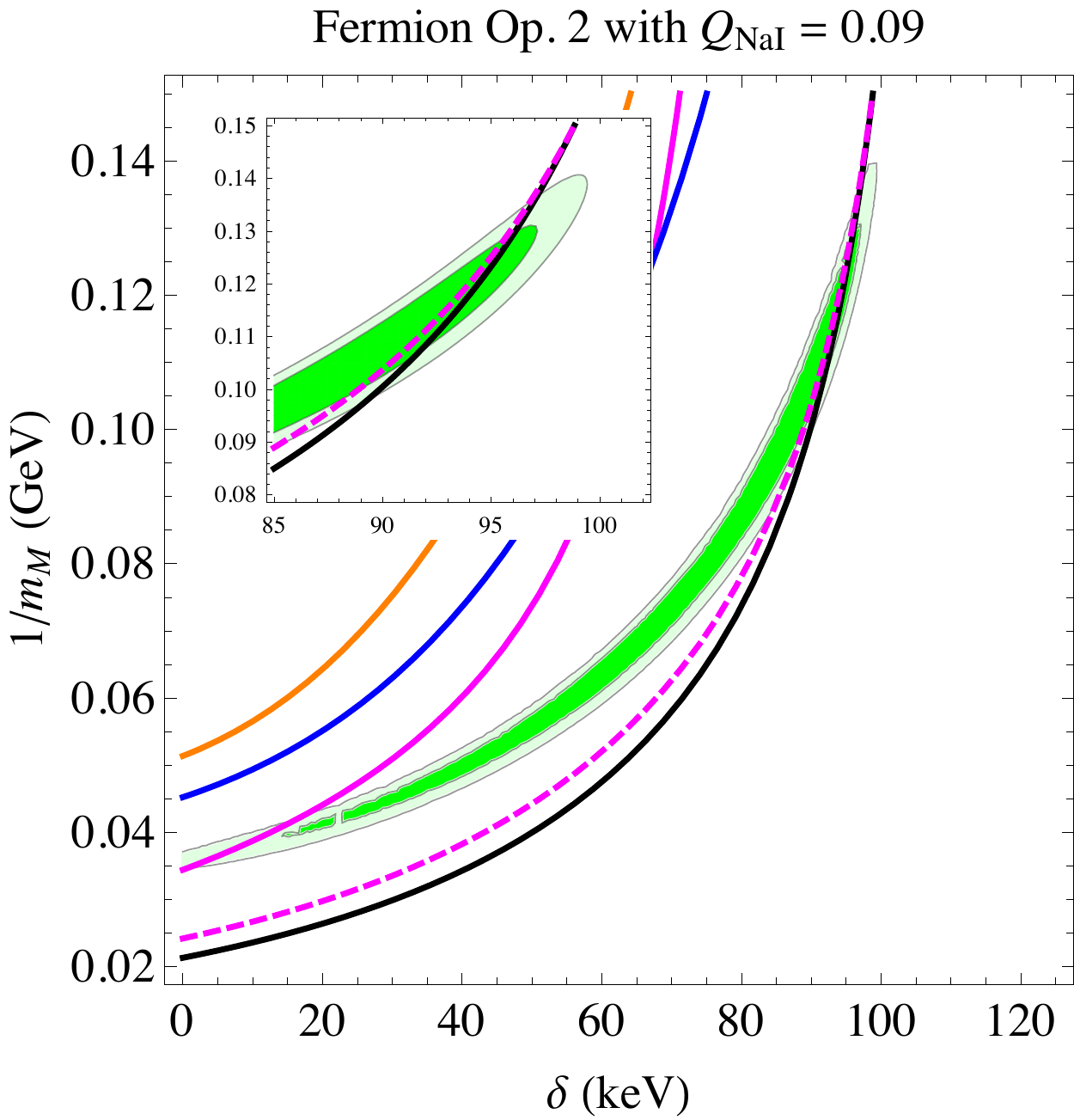}&\includegraphics[scale=0.541]{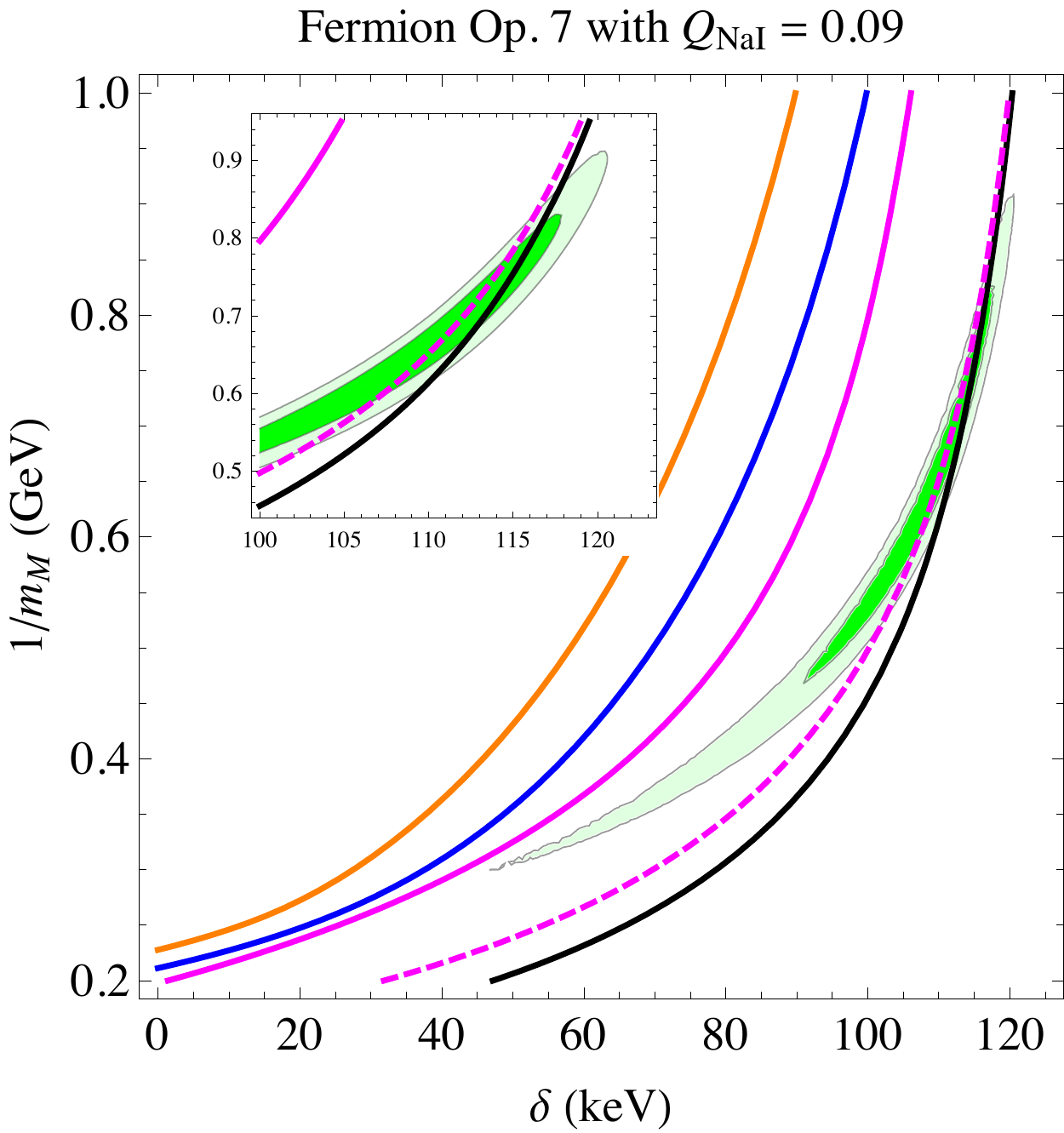}\\
\includegraphics[scale=0.55]{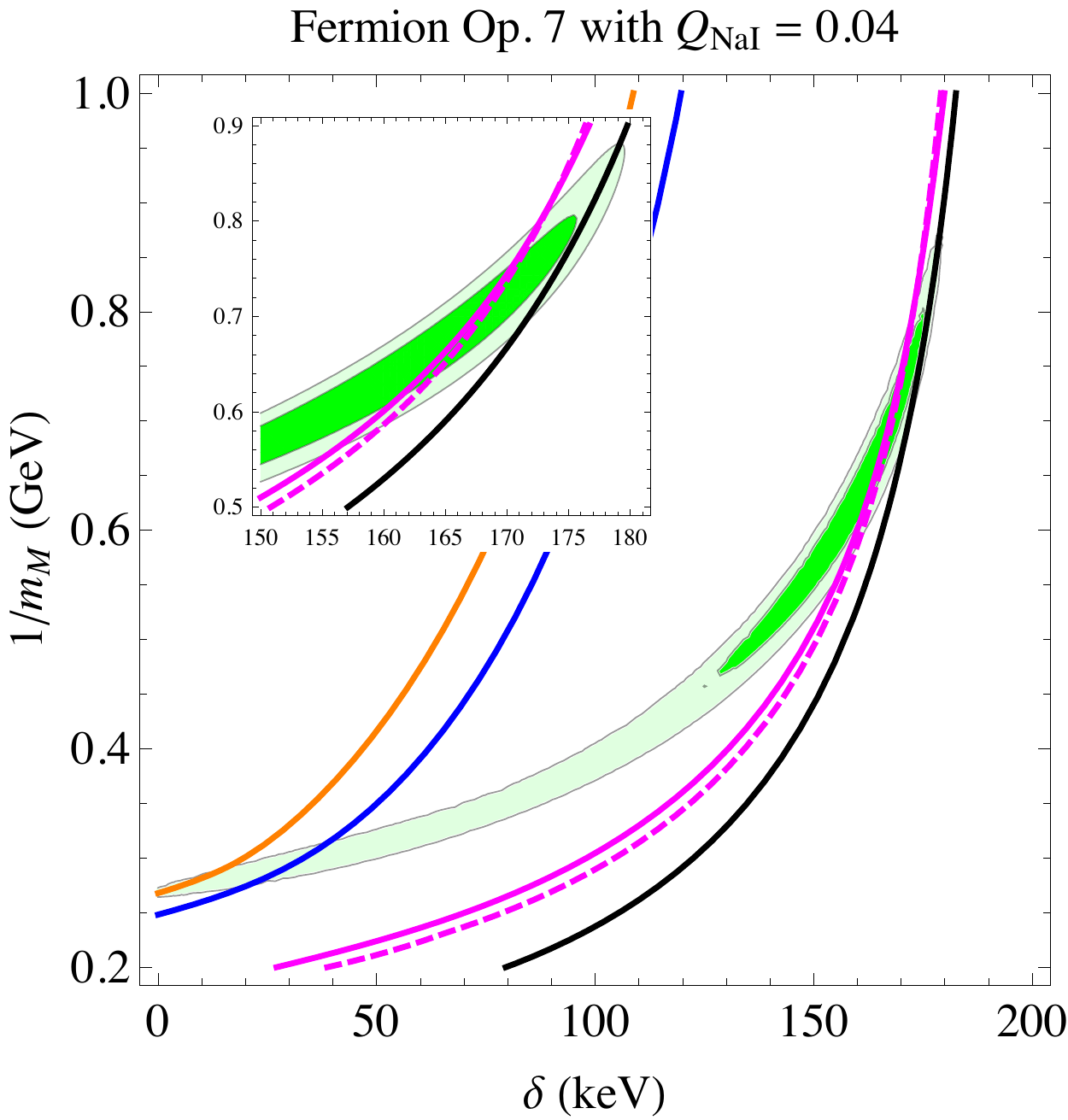}& \includegraphics[scale=0.565]{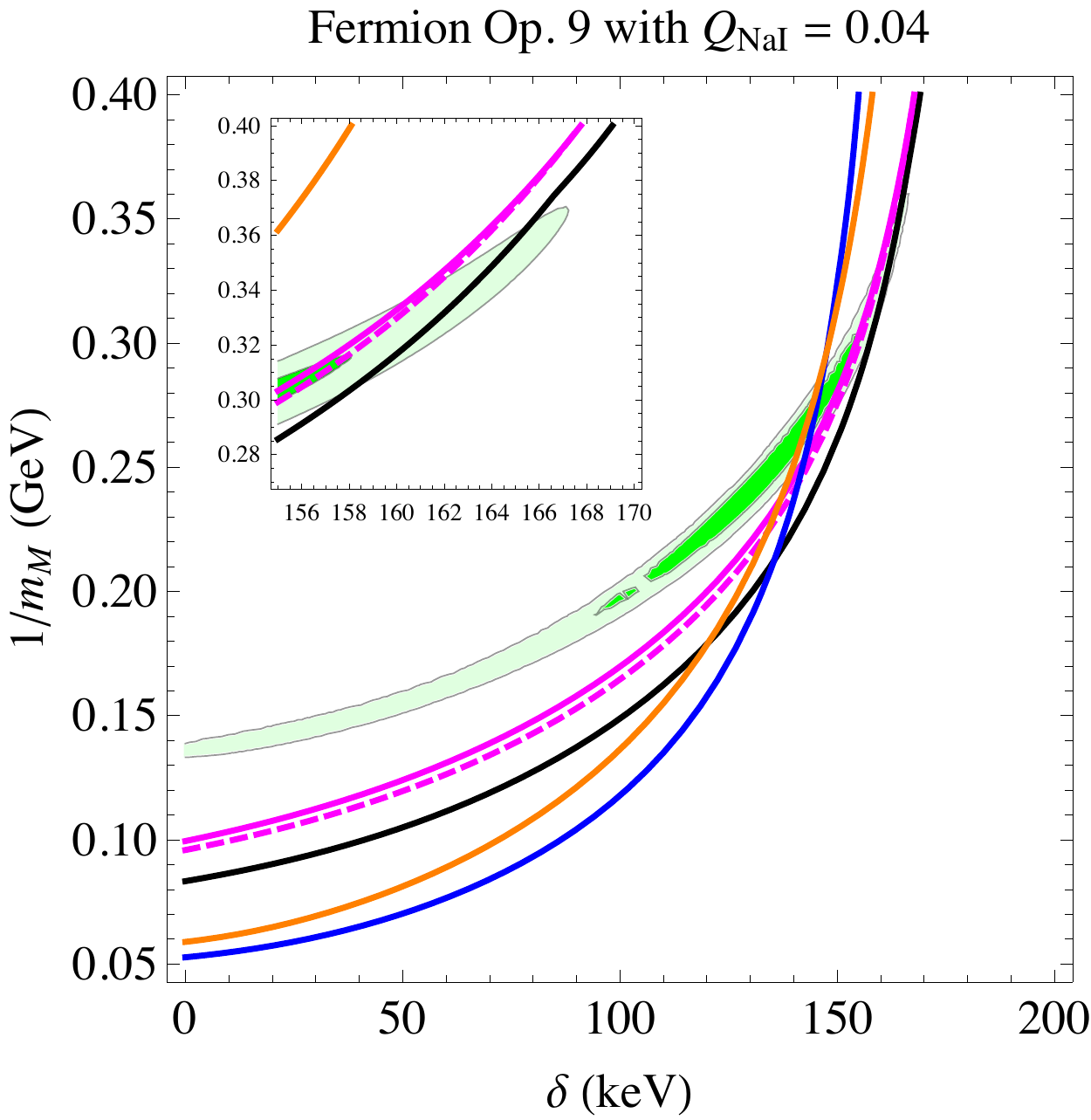}\\
\includegraphics[scale=0.55]{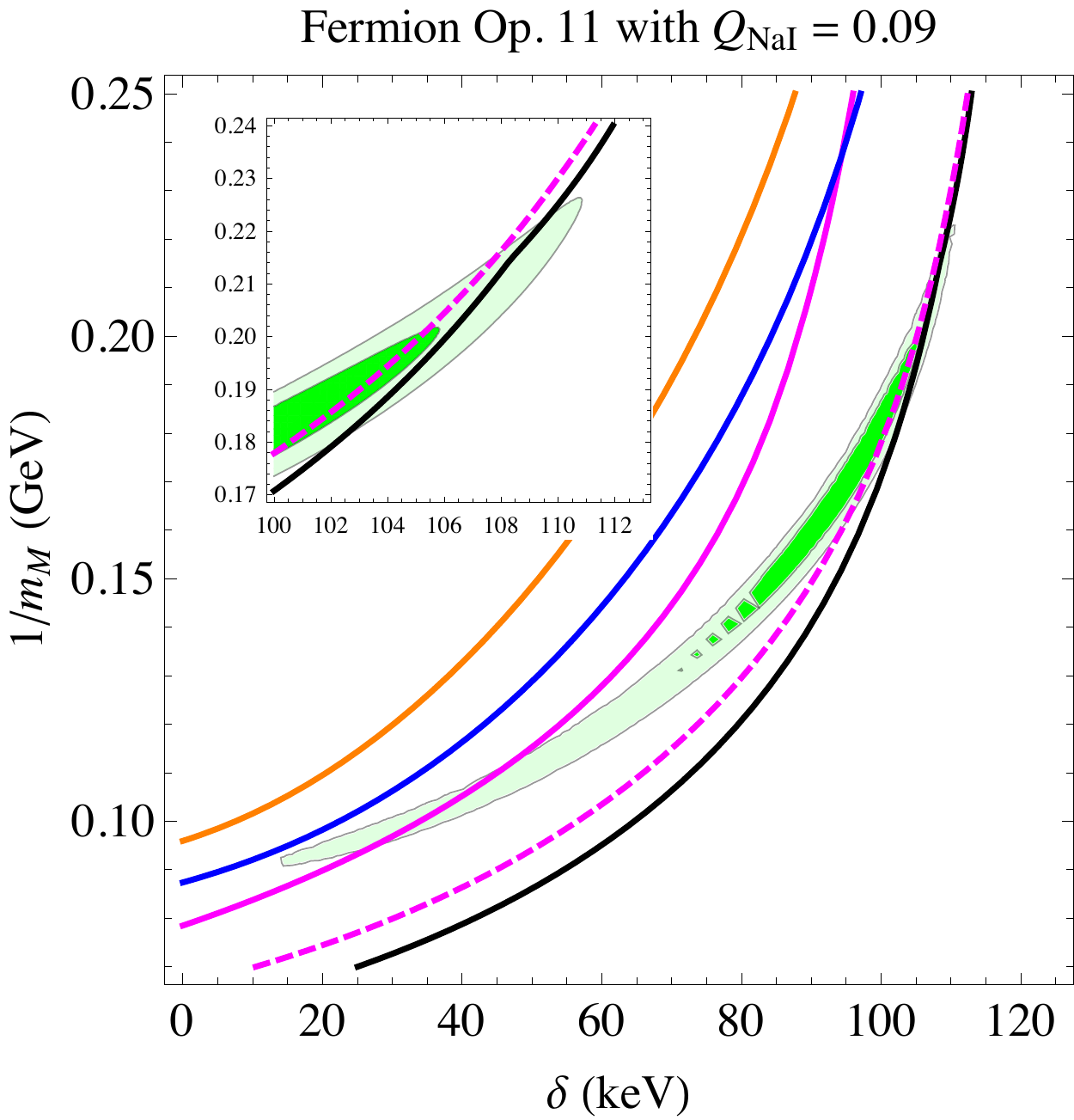}&\includegraphics[scale=0.555]{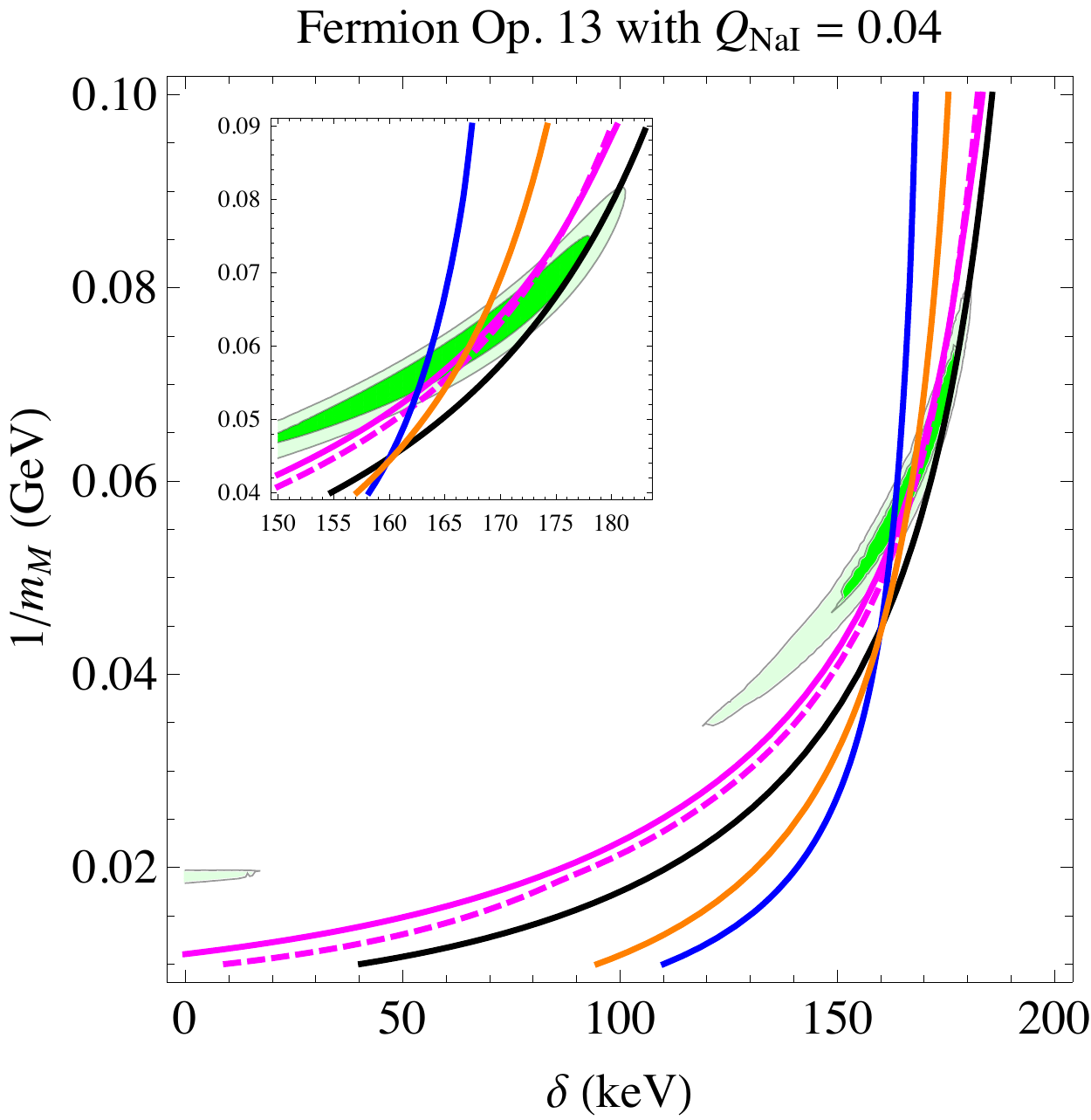}\\
 \end{tabular}
 \renewcommand{\arraystretch}{1.0}
 \end{footnotesize}
 \caption{This figure shows the combined limits plots for operators which have an unconstrained region that fits the DAMA signal. The DM masses used are those listed with the corresponding operator in Table~\ref{tbl:datafermion}. Constraints from LUX (blue), XENON100 (orange), KIMS ($Q_\text{CsI}=0.05$ magenta solid, $Q_\text{CsI}=0.10$ magenta dashed) and COUPP (black) are also shown, with the $90\%$ C.L. limits listed in section \ref{sec:dama}.}
\label{fig:combcomb1}
 \end{figure*}

\begin{figure*}[ht!]
 \begin{footnotesize}
 \centering
 \renewcommand{\arraystretch}{1.6}
 \begin{tabular}{cc}
\includegraphics[scale=0.55]{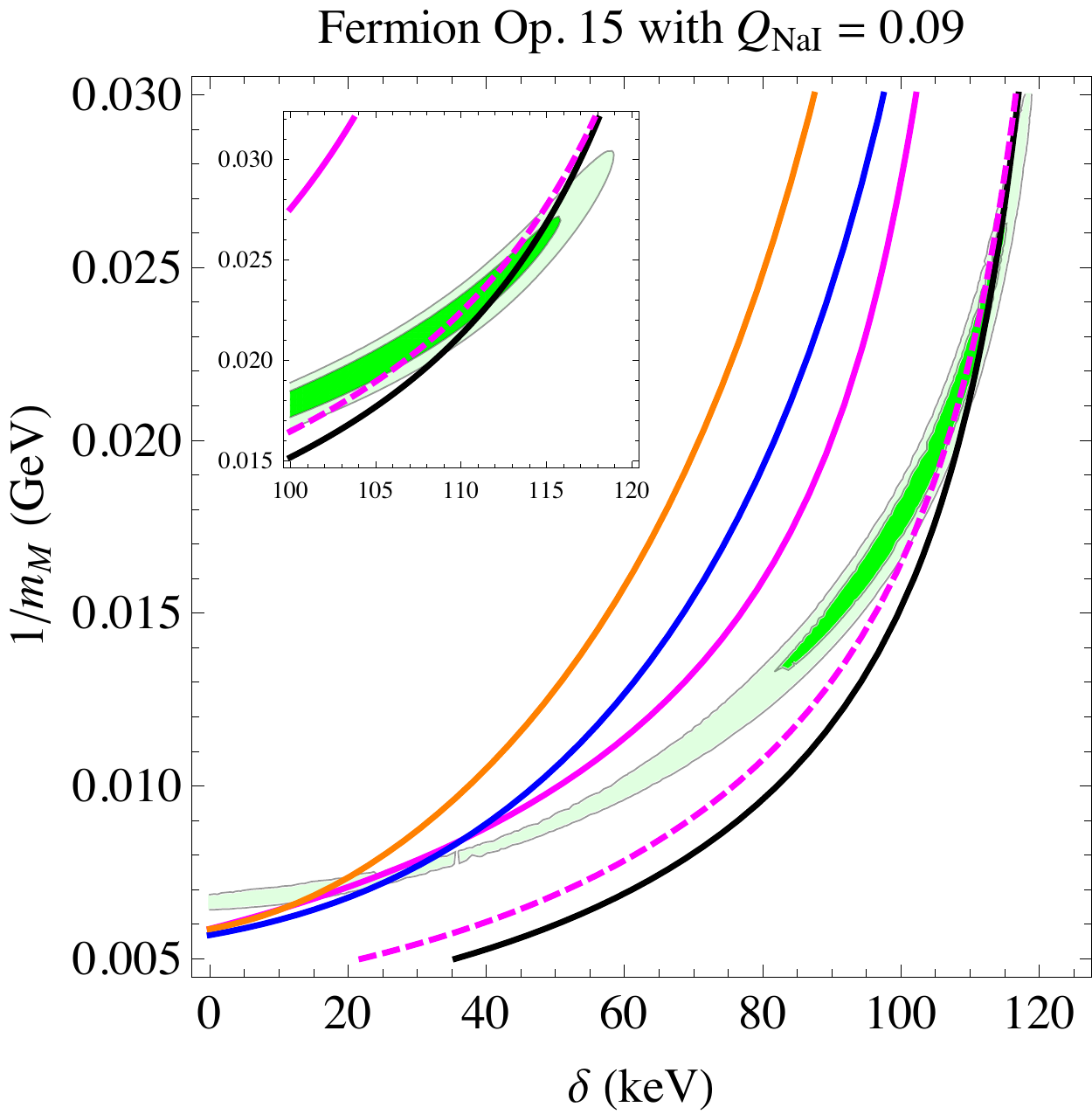}&\includegraphics[scale=0.56]{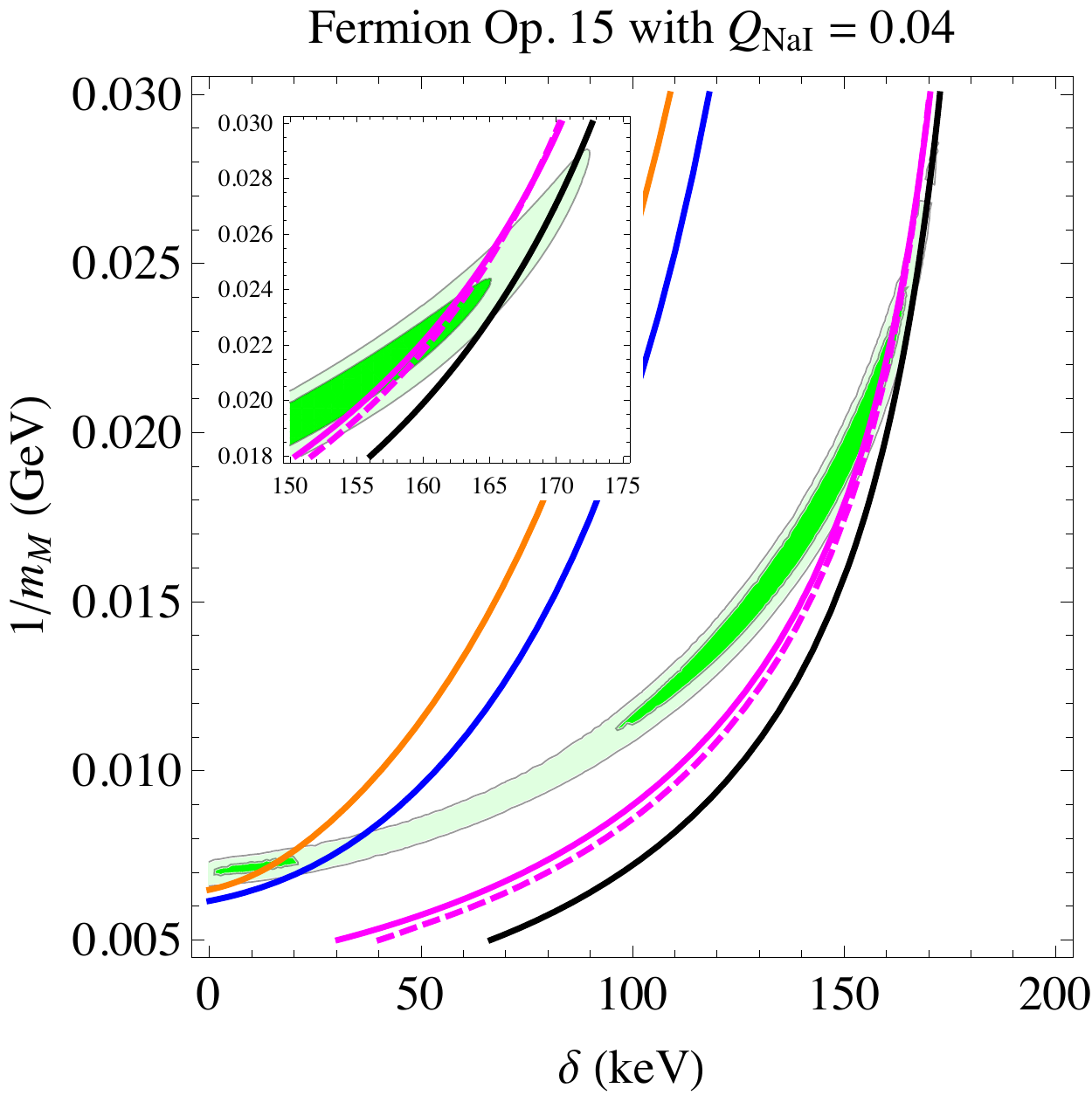}\\\includegraphics[scale=0.55]{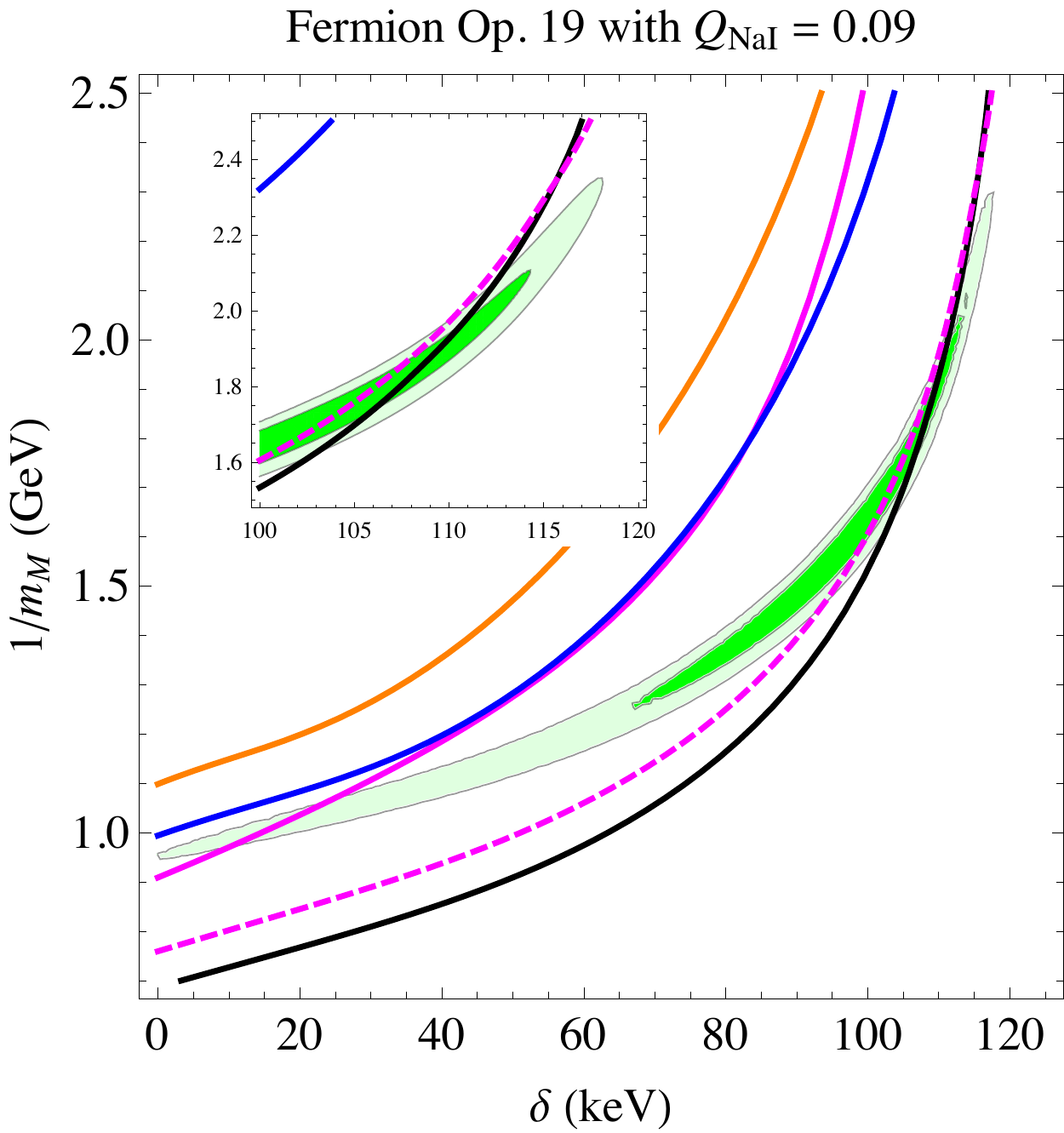}&\includegraphics[scale=0.56]{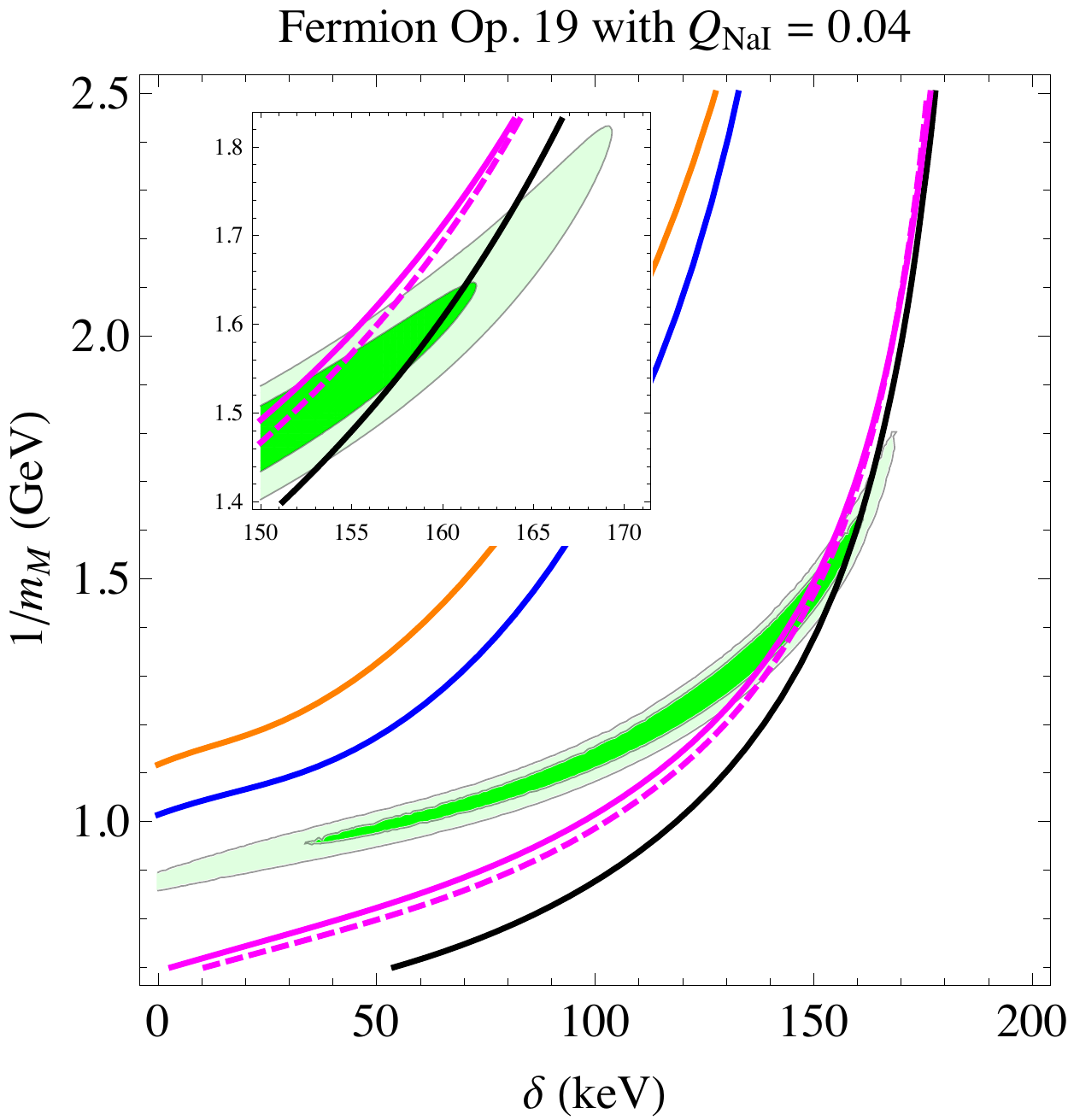}\\
\includegraphics[scale=0.55]{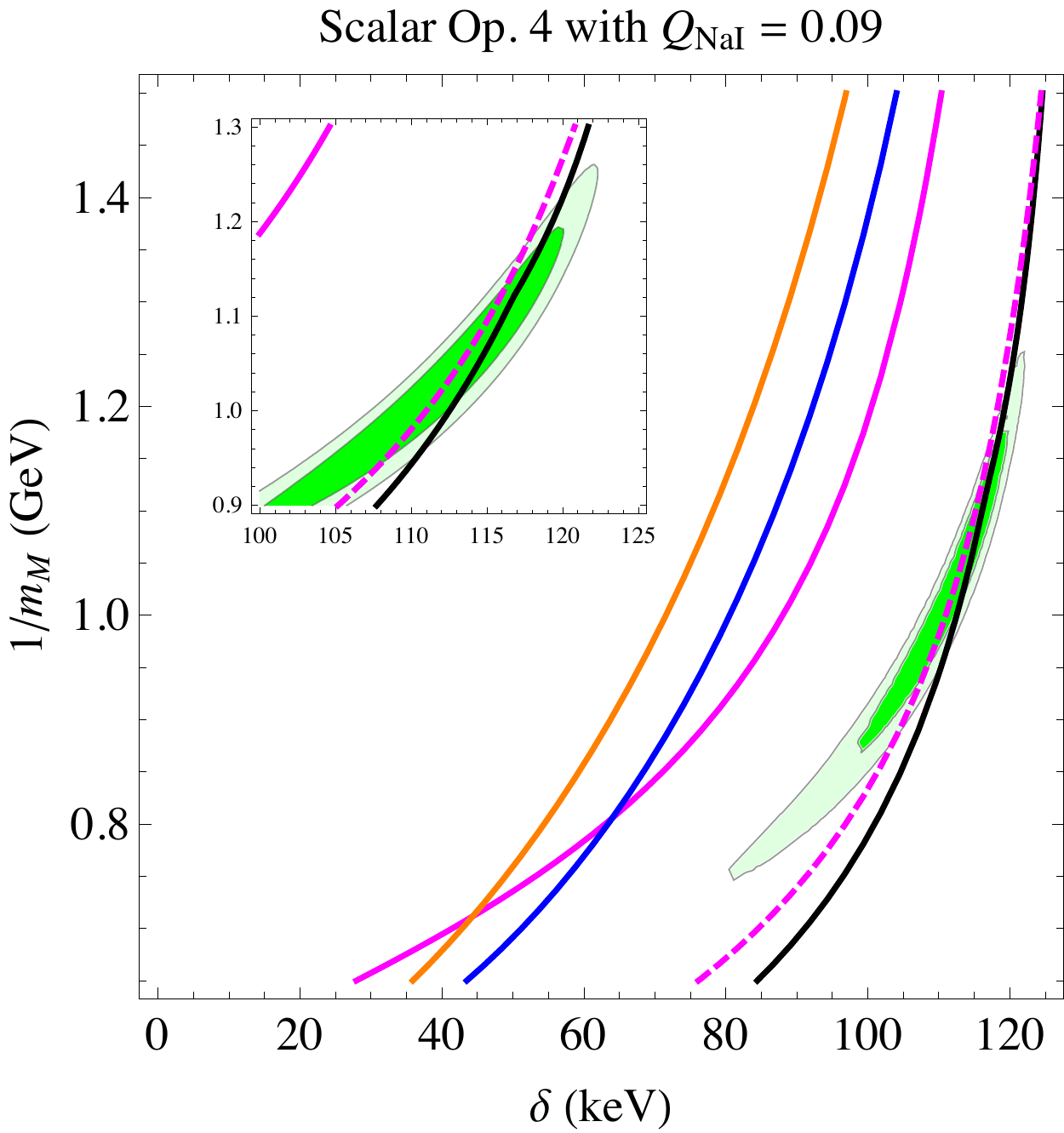}&\includegraphics[scale=0.56] {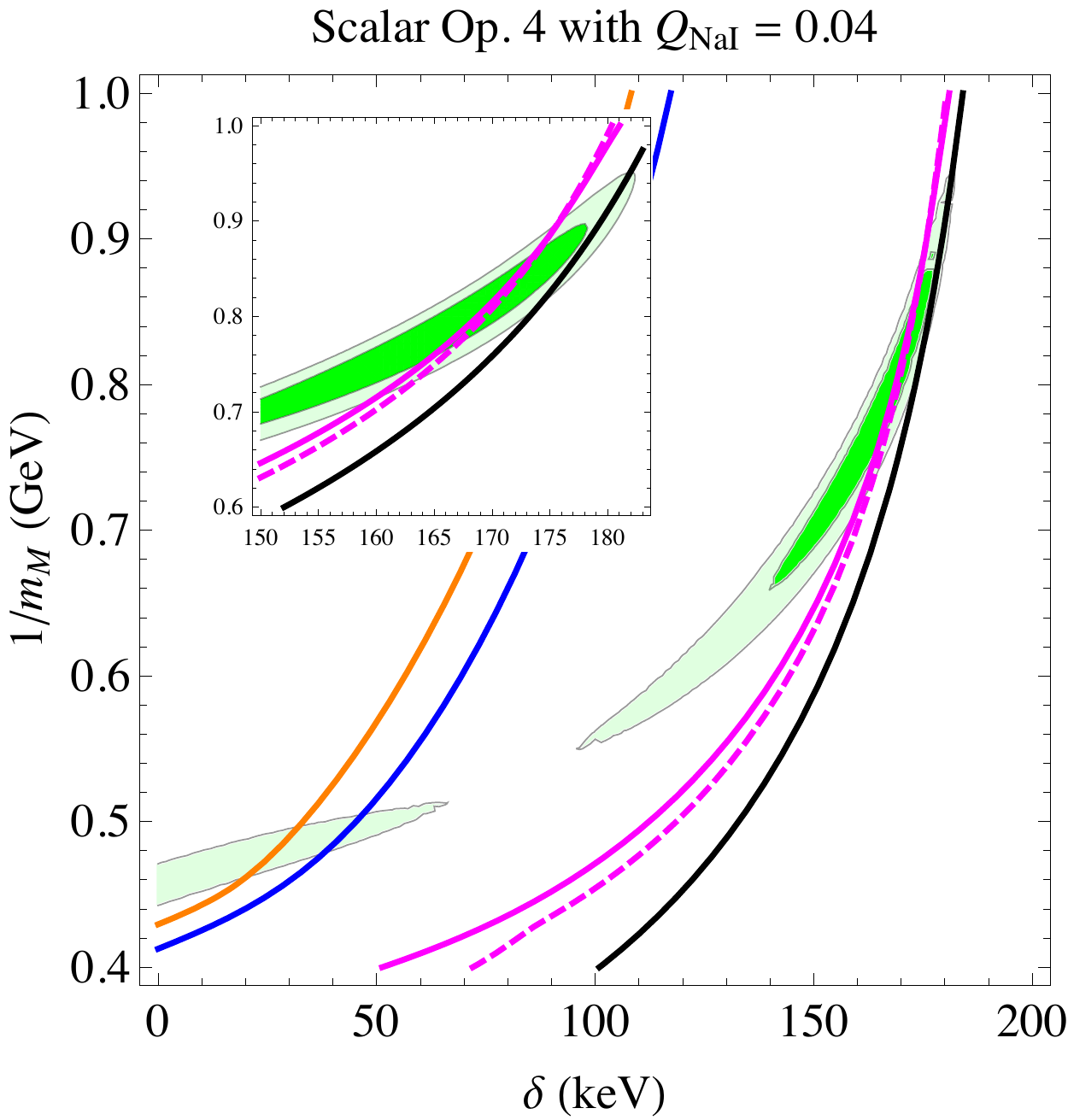}\\
\end{tabular}
 \renewcommand{\arraystretch}{1.0}
 \end{footnotesize}
 \caption{This figure shows the combined limits plots for the remaining operators which have an unconstrained region that fits the DAMA signal.  The DM masses used are those listed with the corresponding operator in Tables \ref{tbl:datafermion}, \ref{tbl:databoson}. Constraints from LUX (blue), XENON100 (orange), KIMS ($Q_\text{CsI}=0.05$ magenta solid, $Q_\text{CsI}=0.10$ magenta dashed) and COUPP (black) are also shown, with the $90\%$ C.L. limits listed in section \ref{sec:dama}.}
\label{fig:combcomb2}
 \end{figure*}

Although the best fit points are ruled out conclusively by COUPP, we still find viable regions of parameter space which maintain a decent fit to DAMA, similar to the case of magnetic inelastic dark matter.  For some of the operators, we found that the DAMA regions could stretch far into the high $\delta$ region of parameter space. The resulting increase in modulation can lead to consistency with the COUPP and KIMS constraints. The fermion operators which have such an allowed region are  operator 2 for $Q_\text{NaI}=0.09$,  operator 7 for both quenching factors, operator 9 for $Q_\text{NaI}=0.04$, 11 for $Q_\text{NaI}=0.09$, 13 for $Q_\text{NaI}=0.04$, 15 with both quenching factors, and 19 with both quenching factors.  Also the scalar to scalar operator 4 has a consistent region for both quenching factors.  For these operators, we have plotted the allowed regions in \figref{fig:combcomb1} and \ref{fig:combcomb2}.    One again can see that the key to avoiding constraints is moving to higher $\delta$.  Thus, the allowed spectra at DAMA will again generically be at slightly higher  energy  with  a slight reduction in the overall amplitude, similar to what was seen in \figref{fig:midmmod}.    In this list of allowed operators, we ignored degeneracies in scattering form factors where we have the families i) fermion 2, fermion 8, scalar 2, and scalar 5, ii) fermion 11, fermion 14, and scalar to vector 7, iii) fermion 15 and scalar to vector 6.  These families share allowed parameter space, although different values for $m_M$ are required to get the same rate.  Interestingly, some operators whose best fit values are only narrowly ruled out remain ruled out in these two dimensional scans.  For instance, fermion operators 4, 10, 20 and scalar to scalar operator 7 have reasonable constraints for $Q_\text{NaI}=0.09$.  In these cases, the form factors do not allow good DAMA fits to persist  to higher $\delta$ thus making it impossible to avoid the constraints.

\section{Conclusions \label{sec:conclusions}}

We have shown that a nonrelativistic effective field theory for the inelastic scattering of dark matter off a nucleus is a straightforward extension of elastic scattering.  The modifications revolve around the Galilean-invariant, incoming dark matter velocity.  Due to the inelastic kinematics, the components of the incident velocity that are perpendicular to the momentum transfer $\vec{q}\;$ have  a new piece that depends on the mass splitting $\delta$
\bea
\vec{v}^\perp_\text{inel} \equiv \vec{v} + \frac{\vec{q}}{2\mu_N} + \frac{\delta}{|\qvec|^2} \vec{q}.
\eea  
This variable change motivates a new basis of scattering matrix elements written in terms of $\vperpinel$.   As an application, we have shown how  inelastic transitions of a fermion to fermion, scalar to scalar, and scalar to vector can be written in terms of this basis.  Finally, since the nuclear matrix elements for most cases only depend linearly on this velocity, we were able to modify the Mathematica code \cite{Anand:2013yka} to generate the form factors for inelastic scattering processes.  Thus, our work extends the framework of \cite{Fitzpatrick:2012ix} so that inelastic dark matter transitions can now be treated in a model independent fashion.  

Armed with our effective field theory, we then created several fits to the DAMA/LIBRA annual modulation.  We considered both the scenario of magnetic inelastic dark matter as well as a model independent survey looking at individual relativistic operators.  Due to the strong constraints from XENON100 and LUX, in the model independent scan, we considered choices for the nucleon couplings that would enhance iodine scattering over xenon.  This led us to consider operators involving only couplings to protons that are sensitive to the proton spin.  CDMS constraints by comparison are significantly weaker due to germanium's lighter mass and even smaller proton spin.     However, we showed that there are significant constraints from the iodine experiments  KIMS and COUPP, which provide a mostly model-independent constraint.  These limits are thus harder to avoid; we find that they can only be weakened by enhanced modulation or by uncertainties in the iodine quenching factors, which affect the KIMS limits. 

For the case of magnetic inelastic dark and for some of the relativistic operators involving only proton couplings, we found that scenarios could be consistent with the DAMA fit and existing constraints.  However, we would like to stress that we are not able to definitively claim a consistent explanation of the DAMA signal.  First of all, due to lack of implementation, we could not treat scattering off of cesium or tungsten, which are relevant for KIMS and CRESST.  Cs in particular has an unpaired proton and should lead to stronger constraints from KIMS.   Hopefully in a future update of the notebook \cite{Anand:2013yka}, these elements could be included.  Second, we only tested the relativistic {\em operators} using our effective field theory.  No models explaining these interactions were considered and thus in a complete model may run into difficulties when confronted with other dark matter constraints.  However, it would be interesting to look at complete models realizing these scenarios, which we leave to future work.  In particular, the magnetic inelastic dark matter scenario should be straightforward to build in a model, since the required coupling structure is through  the standard electromagnetic couplings.    

In the near future, these models should be definitively tested from direct detection experiments alone.  To do so, one high priority is resolving the current uncertainty in iodine quenching factors so as to both pin down the DAMA parameter space and firm up the constraints from KIMS.   Existing data at XENON100 and LUX at energies above 50 keV$_\text{nr}$ should also be reanalyzed which will enhance sensitivities to scenarios when the iodine quenching factor is low.  Finally, iodine target experiments are the most robust tool to rule out or discover these scenarios.  In particular, COUPP's next analysis should give us a definitive answer whether iodine scattering scenarios are a consistent explanation of DAMA's annual modulation signal.    

\section*{Acknowledgements}
We would like to thank G.~Kribs for useful discussions and I.~Yavin for comments on our draft.  We would like to especially thank A.L.~Fitzpatrick for both clarifications of his work and comments on our draft.      
S.C.~was supported in part by the Department of Energy under grant DE-SC0009945 and C.N.~was supported in part by the Department of Energy under grant DE-SC0011640.

%%%%%%%%%%%%%%%%%%%%%%%%%%%%%%%%%%%%%%%%%%%%%%%%%%%%%%%%%%%%%%%%%%%%%%%%%%%%%%%%%%%%%%%%%%%%%%%%%%%%%%%%%%%%%%%%%%%%%%%%%%%%%
%%%%%%%%%%%%%%%%%%			APPENDIX			%%%%%%%%%%%%%%%%%%%%%%%%%%%%%%%%%%%%%%%%%%%%%%%%%%%%%

\appendix

\section{Relativistic Derivation of Nonrelativistic Velocity Operators\label{app:relkinematics}}

As mentioned earlier, there are two ways of constructing the velocity degrees of freedom used in our nonrelativistic field theory: starting with Galilean invariant operators and orthogonalizing them or starting with the relativistic kinematics and reducing to the nonrelativistic limit.  Here we derive the results shown in section \ref{sec:kinematics} using the second method.

To begin, we have the four four-momenta of Fig.~\ref{fig:scattering} from which we need to construct Galilean invariant velocities.  As there are ten constraints; one from energy conservation, three from momentum conservation, four from mass constraints, and two from rotational invariance; we only need two velocity operators.    Using a little foresight, we define three velocities 
\begin{equation}
 \begin{split}
  \vec{v}_N &\equiv \vec{v}_{N_{in}}-\vec{v}_{N_{out}}, \\
  \vec{v}_\chi &\equiv \frac{m_{\chi_1}+m_{\chi_2}}{2m_N}(\vec{v}_{\chi_2}-\vec{v}_{\chi_1}),\text{ and} \\
  \vec v_\text{el}^\perp &\equiv \frac{1}{2}(\vec{v}_{\chi_2}+\vec{v}_{\chi_1}-\vec{v}_{N_{out}}-\vec{v}_{N_{in}}).
 \end{split}
 \label{eq:vel defs}
\end{equation}
and expect to find one relationship between them beyond the orthogonality relations so as to have a total of six degrees of freedom.  The mass factor in front of the relative DM velocity is so that in the elastic limit $\vec v_\chi\rightarrow\vec v_N$.  We also chose the form for $\vec v_\text{el}^\perp$ which is perpendicular to the momentum transfer in the elastic limit and because the velocities have good quantum numbers under $P$, $T$, and hermitian conjugation.

Now that we have our three velocities, we need to orthogonalize them.  We begin with Lorentz invariant combinations:

\begin{eqnarray}
  (p+k)^2 &= &(p'+k')^2 , \nonumber \\
  (p-k')^2 &=& (p'-k)^2 ,\nonumber \\
  k'^2 &=& (p+k-p')^2 ,\text{ and} \label{eq:4mom2 relations for dots} \\
  (p-p')^2 &= &(k-k')^2, \nonumber
\end{eqnarray}

\noindent which we take the nonrelativistic limit of to obtain

\begin{widetext}
\begin{equation}
 \begin{split}
  &-(m_{\chi_1}+m_N)^2-m_{\chi_1} m_N (\vec{v}_{\chi_1}-\vec{v}_{N_{in}})^2 = -(m_{\chi_2}+m_N)^2-m_{\chi_2} m_N (\vec{v}_{\chi_2}-\vec{v}_{N_{out}})^2 , \\
  &-(m_{\chi_1}-m_N)^2+m_{\chi_1} m_N (\vec{v}_{\chi_1}-\vec{v}_{N_{out}})^2 = -(m_{\chi_2}-m_N)^2+m_{\chi_2} m_N (\vec{v}_{\chi_2}-\vec{v}_{N_{in}})^2 , \\
  &-(m_{\chi_1}+m_N-m_{\chi_2})^2-m_{\chi_1} m_N (\vec{v}_{\chi_1}-\vec{v}_{N_{in}})^2+ m_{\chi_1} m_{\chi_2} (\vec{v}_{\chi_1}-\vec{v}_{\chi_2})^2+m_{\chi_2} m_N (\vec{v}_{\chi_2}-\vec{v}_{N_{in}})^2 = -m_N^2 ,\text{ and}\\
  &-(m_{\chi_1}-m_{\chi_2})^2+m_{\chi_1} m_{\chi_2} (\vec{v}_{\chi_1}-\vec{v}_{\chi_2})^2 = m_N^2 (\vec{v}_{N_{in}}-\vec{v}_{N_{out}})^2.
 \end{split}
\end{equation}
\end{widetext}

From these relations we can substitute in the velocities from \eqnref{eq:vel defs} and solve for their dot products.  These are, with the replacement $m_{\chi_2}\rightarrow m_{\chi_1}+\delta$,

\begin{equation}
\begin{split}
 \vec{v}_N\cdot\vec{v}_\chi &= v_\chi^2, \\
 \vec{v}_N\cdot\vec v_\text{el}^\perp &= -\frac{\delta \left((\delta +2 m_{\chi_1})^2+m_N^2 v_\chi^2\right)}{m_N (\delta +2 m_{\chi_1})^2}, \text{ and}\\
 \vec v_\chi\cdot\vec v_\text{el}^\perp &= -\frac{\delta \left((\delta +2 m_{\chi_1})^2 \left(v_N^2+4 (v_\text{el}^\perp)^2+8\right)+4 m_N^2v_\chi^2\right)}{8 m_N (\delta +2 m_{\chi_1})^2}.
\end{split}
\end{equation}

Also, because of the degrees of freedom and our choice of velocities there is a relation between $v_N^2$ and $v_\chi^2$.  This is obtained from the last momentum-conservation equation of \eqnref{eq:4mom2 relations for dots} and is

\begin{equation}
 \frac{4m_{\chi_1}(m_{\chi_1}+\delta)m_N^2}{(2m_{\chi_1}+\delta)^2} v_\chi^2 = \delta^2+m_N^2 v_N^2.
\label{eq:vn2 and vc2 relation}
\end{equation}

The final, orthogonal velocities are given by

\begin{equation}
 \begin{split}
  \vec v_N^\perp &= \vec v_N, \\
  \vec v_\chi^\perp &= \vec v_\chi-\frac{\vec v_\chi\cdot\vec v_N^\perp}{(\vec v_N^\perp)^2}\vec v_N^\perp, \text{ and}\\
  \vec v_\text{inel}^\perp &= \vec v_\text{el}^\perp-\frac{\vec v_\text{el}^\perp\cdot\vec v_N^\perp}{|\vec v_N^\perp|^2}\vec v_N^\perp-\frac{\vec v_\text{el}^\perp\cdot\vec v_\chi^\perp}{|\vec v_\chi^\perp|^2}\vec v_\chi^\perp.
 \end{split}
\end{equation}

As stated in section \ref{sec:kinematics}, we are treating all momenta as order $v$ and $\delta$ as order $v^2$, so the final forms for the velocity operators are, with $\vec v_N\rightarrow\vec q/m_N$,

\begin{equation}
% \begin{split}
  \vec v_N^\perp = \frac{\vec q}{m_N}, 
  \vec v_\chi^\perp = 0, \text{ and }
  \vec v_\text{inel}^\perp = \vec v_\text{el}^\perp +\frac{\delta}{|\vec q\,|^2}\vec q,
% \end{split}
\label{eq:Final orth vel ops}
\end{equation}

\noindent so we only have two velocity-like operators.  As a check, these variables agree with section \ref{sec:kinematics}.

\section{Reduction of Relativistic Operators}

In this paper we have written the nonrelativistic reduction of many relativistic operators, but there are other possibilities not considered here (mainly interactions with spin 2 and beyond mediators).  To help with the reduction of these other operators, we have included a series of reductions for the prototypical elements of a relativistic field theory.  See \cite{Kumar:2013iva} for similar results.  

We concern ourselves with the spinor contractions

\begin{equation}
 \begin{split}
  &\bar\psi_2\psi_1,\quad\bar\psi_2\gamma^5\psi_1,\quad\bar\psi_2\gamma_\mu\gamma^5\psi_1, \\
  &\bar\psi_2\sigma_{\mu\nu}\psi_1,\text{ and}\quad\bar\psi_2\sigma_{\mu\nu}\gamma^5\psi_1,
 \end{split}
 \nonumber
\end{equation}

\noindent where $\sigma_{\mu\nu}\equiv\frac{i}{2}[\gamma_\mu,\gamma_\nu]$.

In the nonrelativistic limit these become

\begin{widetext}
\begin{equation}
 \bar\psi_2\psi_1 \simeq 2\sqrt{m_1}\sqrt{m_2} {\bf 1}_\psi,
 \label{eq:P P}
\end{equation}
\begin{equation}
 \bar\psi_2\gamma^5\psi_1 \simeq 2\sqrt{m_1}\sqrt{m_2} (\vec{v}_1-\vec{v}_2)\cdot \vec{S}_\psi,
 \label{eq:P g5 P}
\end{equation}
\begin{equation}
 \bar\psi_2\gamma_\mu\gamma^5\psi_1 \simeq 2\sqrt{m_1}\sqrt{m_2}(2 S_\psi^i \delta_\mu^i-(\vec{v}_1+\vec{v}_2)\cdot \vec{S}_\psi \delta_\mu^0),
 \label{eq:P gu g5 P}
\end{equation}
\begin{equation}
 \bar\psi_2\sigma_{\mu\nu}\psi_1 \simeq \sqrt{m_1}\sqrt{m_2}\left\{ 4\epsilon_{ijk}S_\psi^k\delta_\mu^i\delta_\nu^j + i(\delta_\mu^0\delta_\nu^a-\delta_\mu^a\delta_\nu^0)\left[-2i\epsilon_{aik}(\vec v_1+\vec v_2)^iS_\psi^k + (\vec v_1-\vec v_2)^a\right] \right\},\text{ and}
 \label{eq:P s P}
\end{equation}
\begin{equation}
 \bar\psi_2\sigma_{\mu\nu}\gamma^5\psi_1 \simeq -\sqrt{m_1}\sqrt{m_2}\left\{ 4i S_\psi^i(\delta_\mu^0\delta_\nu^i-\delta_\mu^i\delta_\nu^0)+\epsilon_{abc}\delta_\mu^a\delta_\nu^b\left[ -2i\epsilon_{cid}(\vec v_1+\vec v_2)^iS_\psi^d + (\vec v_1-\vec v_2)^c \right] \right\}.
 \label{eq:P s g5 P}
\end{equation}
\end{widetext}

In these equations ${\bf 1}_\psi$ is the unit operator in spin-space, $\vec{v}_1$ is the velocity of the incoming $\psi_1$ particle, $\vec{v}_2$ is the velocity of the outgoing $\psi_2$ particle, $\vec{S}_\psi$ is the spin operator for the $\psi$ particle, $g_{\mu\nu}$ is the metric tensor, and $\epsilon_{ijk}$ is the Levi-Civita symbol.  These reductions rely on $\psi_1$  in the initial state and $\psi_2$ in the final state (not their antiparticles) and that the only difference in these particles is the mass ($m_1$ and $m_2$ for initial and final respectively). One can also use the Gordon identity,
\begin{equation}
 \bar\psi_1\gamma_\mu\psi_2 = \frac{1}{2\sqrt{m_1}\sqrt{m_2}}\bar\psi_1(p_{1 \mu}+p_{2 \mu}+i\sigma_{\mu\nu}q^\nu)\psi_2,
\end{equation}
\noindent for the vector interaction.

Another useful result is the nonrelativistic limit for the time-like component of the momentum transfer, which is

\begin{equation}
 \begin{split}
  q^0 &\simeq \delta + \frac{m_{\chi_1}}{2}(\vec v_{\chi_2}^2 - \vec v_{\chi_1}^2), \text{ or} \\
  q^0 &\simeq \frac{m_N}{2}(\vec v_{N_{in}}^2 - \vec v_{N_{out}}^2).
 \end{split}
\end{equation}

\noindent These relations are sometimes needed for the preservation of Galilean invariance but can be easy to overlook.

To reduce operators for spin 1 particles we must take into account the polarization of a nonrelativistic vector boson. This is given by 
\begin{equation}
 \begin{split}
  \varepsilon^0_\lambda(\vec{p}) &\simeq \frac{\vec{p}}{m}\cdot\vec\varepsilon_\lambda(\vec{0}) \\
  \vec\varepsilon_\lambda(\vec{p}) &\simeq \vec\varepsilon_\lambda(\vec{0}),
 \end{split}
 \label{eq:pol trans eqs}
\end{equation}

\noindent to lowest order in $\vec p$.

\section{Transition Amplitude in Nuclear Response Basis}
\label{app:Trans Amp}
Since the effective field theory for inelastic dark matter is so similar to the effective field theory for elastic dark matter, it can be easy to overlook some of the important differences.  The change in the Galilean-invariant incoming dark matter velocity is stressed above, but the possible complex nature for the coefficients of the nonrelativistic operators \eqnref{eqn:operators} is another modification.  To highlight both of these effects we reproduce the relevant results for the squared matrix element, following \cite{Anand:2013yka}.

First we write our Lagrangian as
\begin{equation}
 \mathcal{L}=\sum_{\tau=0,1}\sum_{i=1}^{15}c_i^\tau\mathcal{O}_i,
\end{equation}
where $\tau$ characterizes the isospin structure of the coupling, allowing different couplings to protons and neutrons.  We then calculate the transition amplitude, by  averaging over initial spins and summing over outgoing spins, and expand in the basis of the nuclear responses, giving
%%%
\begin{widetext}
\begin{equation}
 \begin{split}
  &\frac{1}{2j_\chi+1}\frac{1}{2j_N+1}\sum_\text{spins}|\mathcal{M}|^2_\text{nuclear} = \frac{4\pi}{2j_N+1}\sum_{\tau=0,1}\sum_{\tau'=0,1}\Bigg\{ \\
  &\sum_{J=0,2,...}^\infty \bigg[ R_M^{\tau\tau'}(\vinelsq,\qsqmN)\langle j_N||M_{J;\tau}(q)||j_N\rangle\langle j_N||M_{J;\tau'}(q)||j_N\rangle \\
  &\quad\quad+\qsqmN R_{\Phi''}^{\tau\tau'}(\vinelsq,\qsqmN)\langle j_N||\Phi_{J;\tau}''(q)||j_N\rangle\langle j_N||\Phi_{J;\tau'}''(q)||j_N\rangle \\
  &\quad\quad+\qsqmN R_{\Phi''M}^{\tau\tau'}(\vinelsq,\qsqmN)\langle j_N||\Phi_{J;\tau}''(q)||j_N\rangle\langle j_N||M_{J;\tau'}(q)||j_N\rangle\bigg]\\
  &+\sum_{J=2,4,...}^\infty\left[\qsqmN R_{\tilde\Phi'}^{\tau\tau'}(\vinelsq,\qsqmN)\langle j_N||\tilde\Phi_{J;\tau}'(q)||j_N\rangle\langle j_N||\tilde\Phi_{J;\tau'}'(q)||j_N\rangle\right] \\
  &+\sum_{J=1,3,...}^\infty\bigg[ R_{\Sigma''}^{\tau\tau'}(\vinelsq,\qsqmN)\langle j_N||\Sigma_{J;\tau}''(q)||j_N\rangle\langle j_N||\Sigma_{J;\tau'}''(q)||j_N\rangle \\
  &\quad\quad+R_{\Sigma'}^{\tau\tau'}(\vinelsq,\qsqmN)\langle j_N||\Sigma_{J;\tau}'(q)||j_N\rangle\langle j_N||\Sigma_{J;\tau'}'(q)||j_N\rangle \\
  &\quad\quad+\qsqmN R_{\Delta}^{\tau\tau'}(\vinelsq,\qsqmN)\langle j_N||\Delta_{J;\tau}(q)||j_N\rangle\langle j_N||\Delta_{J;\tau'}(q)||j_N\rangle \\
  &\quad\quad+\qsqmN R_{\Delta\Sigma'}^{\tau\tau'}(\vinelsq,\qsqmN)\langle j_N||\Delta_{J;\tau}(q)||j_N\rangle\langle j_N||\Sigma_{J;\tau'}'(q)||j_N\rangle \bigg] \Bigg\}.
 \end{split}
\end{equation}
\end{widetext}
%%%
This result is expanded in spherical harmonics leading to the nuclear operators $M, \Delta, \Sigma', \Sigma'', \tilde{\Phi}', \Phi''$.  The inelastic kinematics does not modify these operators, so we do not reproduce their expressions.   Instead, the changes are solely in the $R$ coefficients
%%%
\begin{widetext}
\begin{equation}
 \begin{split}
  R^{\tau\tau'}_M(|\vec v_T\,|^2,\qsqmN,\delta) &= c_1^\tau c_1^{\tau'*}+\frac{j_\chi(j_\chi+1)}{3}\left[\left(\qsqmN c_5^\tau c_5^{\tau'*}+c_8^\tau c_8^{\tau'*}\right)\vinsqExp+\qsqmN c_{11}^\tau c_{11}^{\tau'*}\right]\\
  R^{\tau\tau'}_{\Phi''}(|\vec v_T\,|^2,\qsqmN,\delta) &= \frac{1}{4}\qsqmN c_3^\tau c_3^{\tau'*}+\frac{j_\chi(j_\chi+1)}{12}\left(c_{12}^\tau-\qsqmN c_{15}^\tau\right)\left(c_{12}^{\tau'*}-\qsqmN c_{15}^{\tau'*}\right)\\    
  R^{\tau\tau'}_{\Phi''M}(|\vec v_T\,|^2,\qsqmN,\delta) &= \text{Re}\left[c_3^\tau c_1^{\tau'*}+\frac{j_\chi(j_\chi+1)}{3}\left(c_{12}^\tau-\qsqmN c_{15}^\tau\right)c_{11}^{\tau'*}\right]\\    
  R^{\tau\tau'}_{\tilde\Phi'}(|\vec v_T\,|^2,\qsqmN,\delta) &= \frac{j_\chi(j_\chi+1)}{12}\left[c_{12}^\tau c_{12}^{\tau'*}+\qsqmN c_{13}^\tau c_{13}^{\tau'*}\right]\\    
  R^{\tau\tau'}_{\Sigma''}(|\vec v_T\,|^2,\qsqmN,\delta) &= \frac{1}{4}\qsqmN c_{10}^\tau c_{10}^{\tau'*}+\frac{j_\chi(j_\chi+1)}{12}\bigg[c_4^\tau c_4^{\tau'*}  +\qsqmN\left(c_4^\tau c_6^{\tau'*}+c_6^\tau c_4^{\tau'*}\right)\\ &\quad +\frac{|\vec q\,|^4}{m_N^4}c_6^\tau c_6^{\tau'*}+\left(c_{12}^\tau c_{12}^{\tau'*}+\qsqmN c_{13}^\tau c_{13}^{\tau'*}\right)\vinsqExp\bigg]\\    
  R^{\tau\tau'}_{\Sigma'}(|\vec v_T\,|^2,\qsqmN,\delta) &= \frac{1}{8}\left[ \qsqmN c_3^\tau c_3^{\tau'*}+c_7^\tau c_7^{\tau'*}\right]\vinsqExp+\frac{j_\chi(j_\chi+1)}{12}\bigg\{c_4^\tau c_4^{\tau'*} \\
        &\quad+\qsqmN c_9^\tau c_9^{\tau'*}+\frac{1}{2}\left[\left(c_{12}^\tau-\qsqmN c_{15}^\tau\right)\left(c_{12}^{\tau'*}-\qsqmN c_{15}^{\tau'*}\right)+\qsqmN c_{14}^\tau c_{14}^{\tau'*}\right]\vinsqExp\bigg\}\\    
  R^{\tau\tau'}_{\Delta}(|\vec v_T\,|^2,\qsqmN,\delta) &= \frac{j_\chi(j_\chi+1)}{3}\left[\qsqmN c_5^\tau c_5^{\tau'*}+c_8^\tau c_8^{\tau'*}\right]\\    
  R^{\tau\tau'}_{\Delta\Sigma'}(|\vec v_T\,|^2,\qsqmN,\delta) &= \frac{j_\chi(j_\chi+1)}{3}\text{Re}\left[c_5^\tau c_4^{\tau'*}-c_8^\tau c_9^{\tau'*}\right].
 \end{split}
\end{equation}
\end{widetext}
%%%
Here we have expanded $\vinelsq$ as in \eqnref{eqn:vinelsq expanded} to show the dependence on $\delta$, and we have also included the appropriate complex conjugation of the coefficients as relativistic inelastic dark matter operators can produce complex coefficients for their nonrelativistic counterparts.

\bibliography{IDM_EFT}

\end{document}